%% file: ArxivSingV2.tex
\documentclass[11pt]{article}
\usepackage{geometry}
\usepackage{fullpage}

\usepackage{amsmath}
\usepackage{amsfonts}
\usepackage{amsthm}
\usepackage[title]{appendix}  
\usepackage{epsfig}            
\usepackage[authoryear, sort]{natbib}
\usepackage{graphicx}
\usepackage{caption}
\usepackage{color, colortbl, xcolor}
\usepackage{multirow}
\usepackage{authblk}
\usepackage[colorlinks,linkcolor=blue,citecolor=blue,urlcolor=blue]{hyperref}
\usepackage[utf8]{inputenc}

\usepackage[ruled,vlined,linesnumbered]{algorithm2e}
\SetKwInput{KwInput}{Input}

\input{Notation.tex}

\title{\LARGE \bf Simultaneous Non-Gaussian Component Analysis (SING) for Data Integration in Neuroimaging}

\author{Benjamin Risk\thanks{Department of Biostatistics and Bioinformatics, Emory University, benjamin.risk@emory.edu} $\mbox{ }$ and
Irina Gaynanova\thanks{Department of Statistics, Texas A\&M University, irinag@stat.tamu.edu}}

\date{}

\newcommand\blfootnote[1]{%
  \begingroup
  \renewcommand\thefootnote{}\footnote{#1}%
  \addtocounter{footnote}{-1}%
  \endgroup
}

\begin{document}

\maketitle

\begin{abstract}
As advances in technology allow the acquisition of complementary information, it is increasingly common for scientific studies to collect multiple datasets. Large-scale neuroimaging studies often include multiple modalities (e.g., task functional MRI, resting-state fMRI, diffusion MRI, and/or structural MRI), with the aim to understand the relationships between datasets. In this study, we seek to understand whether regions of the brain activated in a working memory task relate to resting-state correlations. In neuroimaging, a popular approach uses principal component analysis for dimension reduction prior to canonical correlation analysis with joint independent component analysis, but this may discard biological features with low variance and/or spuriously associate structure unique to a dataset with joint structure. We introduce Simultaneous Non-Gaussian component analysis (SING) in which dimension reduction and feature extraction are achieved simultaneously, and shared information is captured via subject scores. We apply our method to a working memory task and resting-state correlations from the Human Connectome Project. We find joint structure as evident from joint scores whose loadings highlight resting-state correlations involving regions associated with working memory. Moreover, some of the subject scores are related to fluid intelligence.
\end{abstract}
\blfootnote{The authors contributed equally to this work.
}


\section{Introduction}
\label{sec:intro}

\subsection{The importance of data integration in neuroimaging}

Different neuroimaging techniques can provide complementary views of the same information in a person's brain, and integrating this information can provide a more meaningful summary of an individual's brain function. Neuroimaging studies commonly collect multiple modalities from the same set of subjects, where we define modality broadly to include task functional magnetic resonance imaging (fMRI), resting-state fMRI (rs-fMRI), diffusion MRI, structural images, electroencephalography, and positron emission tomography, among others.  A motivating principal is that combining information across modalities will lead to a more accurate understanding of underlying biology compared to one modality alone  \citep{calhoun2016multimodal}.  Studies combining information across datasets have been applied to many scientific questions including the neural correlates of cognition \citep{lerman2017multimodal}, brain networks and alcohol use \citep{crespi2019executive}, brain morphology alterations in bipolar disorder \citep{tang2020fusion}, gray and white matter changes in Alzheimer's \citep{ouyang2015simultaneous}, functional alterations in schizophrenia \citep{sui2013three,sui2011discriminating}, and cognitive control in simultaneous EEG and fMRI \citep{hinault2019spatio}. These studies used data fusion methods to define a set of joint components where each component consists of subject scores (a vector in $\R^n$, where $n$ is the number of subjects) and loadings for each dataset (a vector in $\R^{p_k}$, where $p_k$ is the number of variables in the $k$th dataset). For a given component, the subject scores are equal or highly correlated across datasets. The loadings indicate the importance of a variable relative to the other variables, and a larger subject score indicates the vector of loadings is more important in that individual. The subject scores can also be related to external variables such as behavior and disease status. In our application, we extract linked features across two neuroimaging data types. Our method finds linked structure via the subject scores, and in our application, this results in loadings that exhibit spatial correspondence, even though our formulation does not explicitly incorporate spatial information. We argue this is empirical evidence of meaningful extraction of shared information, whereas shared information is less evident when using existing methods.

\subsection{Motivating Dataset}
Our motivating dataset comes from the young adult Human Connectome Project (HCP), which seeks to characterize brain connectivity and function in healthy adults and examine how brain networks and activity compare across individuals \citep{van2012human}. The HCP comprises high-resolution multi-modality data including 60 minutes of rs-fMRI, task fMRI from seven tasks, diffusion MRI data, and behavioral data from over 1000 subjects, enabling a detailed examination of the human brain \citep{glasser2016human,smith2013resting,barch2013function}. In this paper, we focus on the $N$-back working memory task, which is a popular task used in cognition studies that may be related to fluid intelligence \citep{jaeggi2010concurrent}, and rs-fMRI. These datasets were also analyzed in a data integration study on cognition \citep{lerman2017multimodal}. Task fMRI seeks to manipulate brain neural states while a subject is in the scanner in order to create an activation map. Rs-fMRI is used to examine functional connectivity between brain regions when a subject is staring at crosshairs, in which functional connectivity is commonly measured by the Pearson correlation between the spontaneous activity of different regions. An important scientific goal is to understand whether regions recruited in the performance of a cognitive task are related to spontaneous brain activity. The link between task and rs-fMRI is of particular interest in cognition because it may offer insight into the regional-specialization paradigm, wherein single-brain areas perform specific functions, and the network paradigm of the brain, wherein different parts of the brain interact and may perform multiple functions  \citep{bressler2010large}.

\subsection{Data Integration Methods and Limitations}
Discovering information that is shared between datasets is a fundamental problem in statistics dating back to early work in canonical correlation analysis (CCA) \citep{harold1936relations}. 
This problem has received renewed attention in the Omics era, where genetics studies often collect multiple data types, including methylation, gene expression, copy number, and mutation. Combining information across data types has been used to estimate subject scores that cluster cancer subtypes, which may be useful in biomarker development \citep{mo2013pattern,lock2013joint,Gaynanova:2017te}. Methods have been developed to address the novel challenges in genetics datasets, in particular the $p \gg n$ setting, including sparse CCA \citep{witten2009penalized}, joint and individual variation explained (JIVE) \citep{lock2013joint,feng2018angle}, common and individual feature extraction (CIFE) \citep{zhou2015group}, and multi-omics factor analysis \citep{argelaguet2018multi}. For a comprehensive list of methods, see \cite{Love:2019}. However, data integration methods tailored for multimodal neuroimaging have received comparably less attention and must address distinct challenges due to the different statistical properties of imaging data, most notably that imaging features capturing brain activation and intrinsic functional connectivity are highly non-Gaussian, as discussed further in Section \ref{sec:decomp_onedataset}.

Methods using principal component analysis (PCA) for dimension reduction and independent component analysis (ICA) for subspace interpretation have been proposed in the neuroscience literature  \citep{calhoun2016multimodal,sui2012review,zhou2016linked}. ICA is a popular approach in neuroimaging, where computationally scalable objective functions based on marginal non-Gaussianity extract spatial components in task fMRI \citep{sui2010cca+} and derive interesting network structure in resting-state correlations \citep{amico2017mapping}.
In Joint ICA, also called concatenation ICA, data are standardized and concatenated to form a single large dataset, then PCA is applied, and finally the PCs are rotated to estimate joint independent components (ICs) \citep{Calhoun:2006cp, Calhoun:2006gm,Calhoun:2009jr}. This results in joint subject scores that are equal in the two datasets. This method is very flexible in that the datasets have the same subjects but can be of different dimensions. Multimodal CCA with Joint ICA (mCCA+jICA) is a popular extension that first applies PCA separately to each dataset to reduce the number of features \citep{sui2011discriminating}. It then applies a modified CCA to find highly correlated joint subject scores, and third, it adjusts these scores and loadings using Joint ICA. As noted above, it was used to examine activation maps from two tasks ($N$-back and relational processing), resting-state correlations, and cortical thickness in \cite{lerman2017multimodal}, and subject scores were related to cognitive measures.
Parallel ICA is a method that iterates between ICA for the separate datasets and maximizing the correlation between matched subject scores \citep{liu2009combining,vergara2014three}. Linked ICA uses a variational Bayes approximation to estimate shared subject scores, and it allows the scaling of the scores to vary between datasets \citep{groves2011linked}. There are also joint and individual methods for independent components in a single modality where structure is shared in the voxel dimension \citep{pakravan2018extraction}, but here we focus on the multimodal integration studies. Although the maximization of non-Gaussianity used in the ICA steps results in components that better separate brain regions than PCA-only methods, these methods have other limitations. We will argue that 1) previous methods can miss joint structure, for example when PCA discards low variance but biologically important components, and 2) previous methods can spuriously associate structure unique to a dataset with joint structure.

\subsection{Goal of this paper}
Although Joint ICA, mCCA+jICA, and related methods have been widely used in the neuroimaging community, novel approaches that utilize non-Gaussianity for both dimension reduction and latent variable extraction may offer new insights. Projection pursuit and non-Gaussian component analysis (NGCA) have been used to find low-rank structure based on maximizing non-Gaussian measures of information, which contrasts with PCA based on maximizing variance \citep{bickel2018projection,friedman1974projection,blanchard2005non,risk2017linear,virta2016projection,nordhausen2017asymptotic}. However, projection pursuit and NGCA have not been extended to multimodal data analyses. Additionally, the likelihood in the LNGCA model in \cite{risk2017linear} can not be applied to $p \gg n$ due to a rank-deficient $p \times p$ covariance matrix. Characterizing shared structure requires a novel model that applies to $p \gg n$, which in turns requires a new estimation framework. We propose Simultaneous Non-Gaussian component analysis (SING), which formulates an objective function based on maximizing the skewness and kurtosis of latent components with a penalty to encourage similarity between subject scores. Similar to JIVE \citep{lock2013joint}, our approach focuses on information shared in subject score subspaces.   
Subject scores can be viewed as weighing the importance of the corresponding non-Gaussian components in the subjects' datasets, and our decomposition allows the scaling to differ between datasets.  

In summary, our contributions are the following:
\begin{enumerate}
    \item  We propose a new matrix decomposition for shared non-Gaussian structure across datasets, and we derive an estimation algorithm based on the use of higher-order moments.
    \item The proposed approach improves estimation of subject scores compared to popular methods for multimodal analysis in neuroimaging.
    \item In the analysis of the HCP data, the joint loadings in SING capture spatially localized patches of contiguous cortex in the working memory task and the spatially coinciding node in the rs correlations along with its edges, while joint structure is less apparent in the existing approaches. Additionally, we find the joint subject scores of a subset of the components are significantly associated with fluid intelligence. 
\end{enumerate}

The remainder of this paper is organized as follows. In Section 2, we propose the novel model and algorithm called SING. In Section 3, we compare SING to Joint ICA and mCCA+jICA, and we evaluate the impact of the penalty controlling the similarity between subject scores in SING. In Section 4, we analyze working-memory task activation maps and resting-state correlations from the HCP. Section 5 presents a summary and discussion. 

\section{Methods}
\label{sec:meth}

\subsection{Matrix decomposition for one dataset}\label{sec:decomp_onedataset}
We first summarize a matrix decomposition for a single dataset $\bX \in \R^{n \times p_x}$ ($n$ subjects and $p_x$ features) into a non-Gaussian subspace and a Gaussian subspace based on linear non-Gaussian component analysis (LNGCA). This matrix decomposition differs from \citep{risk2017linear}, who used a likelihood with statistical independence between components, whereas our matrix decomposition uses non-Gaussianity with orthogonality constraints. This is discussed further in Remark 2.

Each row of $\bX$ is a vector of features from the $i$th subject. Let $\bX_c$ denote the double-centered data matrix such that $\bone^\top \bX_c = \bzero^\top$ and $\bX_c \bone = \bzero$, which has rank $n-1$ when $p_x > n$. Let $\bI_{r_x}$ denote the $r_x \times r_x$ identity matrix. Then define the decomposition
\begin{align}
\bX_c = \bM_x \bS_x + \bM_{Nx} \bN_x,\label{eq:modellngca}
\end{align}
where
\begin{enumerate}
\item $\bM_x \in \tR^{n \times r_x}$ and $\bM_{Nx} \in \tR^{n \times (n-r_x-1)}$. The columns of $\bM_x$ are called subject scores, and the matrix $[\bM_x, \bM_{Nx}]$ is called the mixing matrix and has rank $n-1$.   
\item $\bS_x \in \tR^{r \times p_x}$ and $\bN_x \in \tR^{(n-r_x-1) \times p_x}$. $\bS_x \bS_x^\top = p_x \bI_{r_x}$, $\bN_x \bS_x^\top = \bzero_{(n - r_x -1) \times r_x}$. The rows of $\bS_x$ are the non-Gaussian components, and elements of $\bS_x$ are called variable loadings because $\bX_c \bS_x^\top = \bM_x$. The rows of $\bN_x$ are the Gaussian components.
\item The rows of $\bS_x$ have maximum non-Gaussianity subject to the aforementioned orthogonality constraints.
\end{enumerate}
The goal is to estimate $\bM_x$ and $\bS_x$, while the Gaussian components are regarded as noise. Note that any signed permutation of the columns of $\bM_x$ and corresponding signed permutation of the rows of $\bS_x$ also satisfies \eqref{eq:modellngca}. The double centering results in mean-zero scores and mean-zero components, which is useful for defining shared structure in Section \ref{sec:2.2}. 

The decomposition~\eqref{eq:modellngca} is fitted based on maximizing the non-Gaussianity of the features. This is useful in neuroimaging because 1) the goal is to maximize the non-Gaussian measure in the image space, where vectorized features like brain activation maps and resting-state networks have highly non-Gaussian distributions; and 2) $p \gg n$, where in general, the number of subjects is  small relative to the number of voxels or edges. Maximizing non-Gaussianity across voxels is also recommended in multi-subject spatial ICA of fMRI, which has been used in over 20,000 papers \citep{calhoun2012multisubject}, and is used in ICA decompositions of vectorized correlation matrices \citep{amico2017mapping}.  Maximizing non-Gaussianity across edges and spatial locations is used in mCCA+jICA \citep{lerman2017multimodal}, where the ICA step results in improved separation of sources compared to mCCA \citep{sui2011discriminating}. We note that from a probabilistic perspective, this corresponds to a model in which the number of subjects is fixed, which will be discussed in Section \ref{sec:model}.

To obtain the decomposition in \eqref{eq:modellngca}, we must find an unmixing matrix $\bA_x$ such that $\bA_x \bX_c \bX_c^\top \bA_x^\top = \bS_x \bS_x^\top = p_x \bI_{r_x}$. To satisfy this constraint, we reparameterize the model using a pre-whitening matrix. Let $\widehat \bSigma_x = \bX_c \bX_c^\top/p_x$. Then define the eigenvalue decomposition $\widehat \bSigma_x = \bV_x \bLambda_x \bV_x^\top$ and the pre-whitening matrix $\hatbL_x = \bV_x \bLambda_x^{-1/2} \bV_x^\top$. This is defined for rank-degenerate $\bX_c$ via the economy eigenvalue decomposition. Note $\bA_x = \bU_x \hatbL_x$. Let $f()$ be a measure of non-Gaussianity.
Then \eqref{eq:modellngca} is estimated using
\begin{equation}\label{eq:LNGCAsep}
\begin{split}
\minimize_{\bU_x}& \;\; -\sum_{l=1}^{r_x}f(\bu_{xl}^{\top} \hatbL_x \bX_c)\\
    \mbox{subject to}&\quad \bU_x\bU_x^{\top} = \bI_{r_x},
\end{split}
\end{equation}
where $\bu_{xl}^\top$ is the $l$th row of the $r_x \times n$ matrix $\bU_x$. 

The key difference between this matrix decomposition and ICA approaches in \cite{Calhoun:2006cp,beckmann2004probabilistic} is that while we retain all $n$ directions of variance during pre-whitening, the latter only retain $r_x \ll n$ directions associated with maximum variance. Specifically, let $\bV_{r_x}$ and $\bLambda_{r_x}$ denote the first $r_x$ eigenvectors and eigenvalues of $\widehat{\bSigma}_x$. Let $\cO_{r_x \times r_x}$ denote the class of $r_x \times r_x$ orthonormal matrices. The ICA objective function estimating $r_x < n$ components is
 \begin{align}
 \underset{\bU_x \in \cO_{r_x \times r_x}}{\argmin} & \;\; -\sum_{l=1}^{r_x}f(\bu_{xl}^{\top} \bLambda_{r_x}^{-1/2} \bV_{r_x}^\top \bX_c),\label{eq:ICA}
 \end{align}
where $\bu_{xl}^\top$ is the $l$th row of $\bU_x$. Some fMRI ICA implementations whiten using $(\bLambda_{r_x} - \sigma^2 \bI_{r_x})^{-1/2} \bV_{r_x}^\top$ where $\sigma^2$ is the average of the eigenvalues of the discarded directions \citep{beckmann2004probabilistic}. Let $\hatbS_x^{ICA}$ denote the components resulting from \eqref{eq:ICA}. Then $\hatbS_x^{ICA}$ maximizes the non-Gaussianity of the principal component loadings (scaled to have norm equal to one), whereas in LNGCA, $\hatbS_x$ maximizes the non-Gaussianity of $\bX_c$. Since the set of matrices defined by $\bU_{r_x}^{\top}\bLambda_{r_x}^{-1/2} \bV_{r_x}$, $\bU_{r_x} \in \cO_{r_x \times r_x}$ is a subset of $\bU_x \hatbL_x$, $\bU_x \in \cO_{r_x \times n}$, LNGCA achieves greater non-Gaussianity.

We measure non-Gaussianity using the Jarque-Bera (JB) statistic, which is a weighted combination of squared skewness and kurtosis \citep{jarque1987test} and was applied to LNGCA in \cite{virta2016projection}. For a vector $\bs\in \mathbb{R}^p$, the JB statistic is
\begin{equation}\label{eq:JB}
    f(\bs) = 0.8\left(\frac1{p} \sum_j s_j^3\right)^2 + 0.2\left(\frac1{p} \sum_j s_j^4 - 3\right)^2.
\end{equation}
Unlike some measures of non-Gaussianity, such as the logistic function used in Infomax \citep{bell1995information}, the JB statistic extracts both sub- and super-Gaussian components.

\subsection{Matrix decomposition for two datasets}\label{sec:model}

We now propose a matrix decomposition of two datasets $\bX \in \tR^{n \times p_x}$ and $\bY \in \tR^{n \times p_y}$ into a joint non-Gaussian subspace defined by shared subject score directions, individual non-Gaussian subspace (where here individual means unique to a dataset), and a Gaussian subspace. Let $r_J$ denote the rank of the joint non-Gaussian subspace. 
Define the double-centered $\bX_c$ such that $\bone^\top \bX_c = \bzero^\top$ and $\bX_c \bone = \bzero$, and similarly for $\bY$. Let $r_x$ and $r_y$ denote the rank of the non-Gaussian subspace (i.e., signal rank) for datasets $\bX_c$ and $\bY_c$, respectively. We consider
\begin{equation}\label{eq:model}
\begin{split}
 \bX_c &= \bM_{J}\bD_x \bS_{Jx} + \bM_{Ix}\bS_{Ix} + \bM_{Nx}\bN_x,\\
 \bY_c &= \bM_{J}\bD_y \bS_{Jy} + \bM_{Iy}\bS_{Iy} + \bM_{Ny}\bN_y,
\end{split}
\end{equation}
where 
\begin{enumerate}
\item $\bM_J \in \R^{n \times r_J}$, $\bM_{Ix} \in \tR^{n \times (r_x - r_J)}$, $\bM_{Iy} \in \tR^{n \times (r_y - r_J)}$, $\bM_{Nx} \in \tR^{n \times (n - r_x - 1)}$, and $\bM_{Ny} \in \tR^{n \times (n - r_y - 1)}$. Additionally, the columns of $\bM_J$ have unit L2 norm to allow explicit scaling via $\bD_x$ and $\bD_y$. 
\item $\bD_x$ and $\bD_y$ are diagonal and allow the size of the joint signal to vary between datasets.
\item $\bS_{Jx}$ are the joint non-Gaussian components, $\bS_{Ix}$ are the individual non-Gaussian components, and $\bN_x$ are the individual Gaussian components, respectively, for $\bX$, with $\bS_{Jx} \bS_{Jx}^\top = p_x \bI_{r_J}$, $\bS_{Ix}\bS_{Ix}^\top  = p_x \bI_{r_x - r_J}$, $\bS_{Jx}\bS_{Ix}^\top  = \bzero_{r_J \times (r_x - r_J)}$, $\bN_x \bS_{Jx}^\top = \bzero_{(n - r_x-1) \times r_J}$, $\bN_x \bS_{Ix}^\top = \bzero_{(n-r_x-1) \times (r_x - r_J)}$, and similarly define the components of $\bY$. 
\item The rows of $[\bS_{Jx}^\top, \bS_{Ix}^\top]^\top$ and $[\bS_{Jy}^\top,\bS_{Iy}^\top]^\top$ have maximum non-Gaussianity subject to the aforementioned orthogonality constraints and the shared $\bM_J$ defined in \eqref{eq:model}. 
\end{enumerate}
The primary goal is to estimate $\bM_J$, $\bS_{Jx}$, and $\bS_{Jy}$. While the individual components may also be of interest, the Gaussian components are regarded as noise.  
\vspace{0.5cm}

\emph{Remark 1.} We can treat \eqref{eq:model} as a probabilistic model by defining random vectors $\bx \in \tR^n$ and $\by \in \tR^n$. Let $\bs_{Jx}\in \tR^{r_J}$ and $\bs_{Ix} \in \tR^{r_x - r_J}$ be non-Gaussian random vectors with mean zero and variance equal to one, and let $\bn_x \in \tR^{n-r_x-1}$ be standard Gaussian. We further assume $[\bs_{Jx}^\top,\bs_{Ix}^\top,\bn_x^\top]$
have mutually independent elements. Similarly define $[\bs_{Jy}^\top,\bs_{Iy}^\top,\bn_y^\top]$ for $\by$. Then the probabilistic model is
\begin{equation}\label{eq:probmodel}
\begin{split}
    \bx &= \bM_{J}\bD_x \bs_{Jx} + \bM_{Ix}\bs_{Ix} + \bM_{Nx}\bn_{x},\\
 \by &= \bM_{J}\bD_y \bs_{Jy} + \bM_{Iy}\bs_{Iy} + \bM_{Ny}\bn_{y}.
 \end{split}
\end{equation}
Now consider $\bx$. The concatenated matrix $[\bM_J \bD_x, \bM_{Ix}]$ is unique up to signed permutations of the columns from Theorem 10.3.9 in \citep{kagan1973characterization}; see also Theorem 1 in \cite{risk2017linear}. Let $\bM_x = [\bM_J \bD_x, \bM_{Ix}] \bP$ be the mixing matrix for the separate LNGCA model of $\bx$, where $\bP$ belongs to the class of signed permutation matrices $\mathcal{P}_{\pm}$. Similarly define $\bM_y$. Consider the squared chordal distance
\begin{equation}\label{eq:distance}
d(\bx,\by) = \left\|\frac{\bx\bx^{\top}}{\|\bx\|_2^2} - \frac{\by\by^{\top}}{\|\by\|_2^2}\right\|_F^2,
\end{equation}
which is between 0 and 2. Then uniqueness implies that $\bM_J\bD_x$ and $\bM_J\bD_y$ must correspond to the matched columns of $\bM_x$ and $\bM_y$ that have chordal distance equal to zero, where the chordal distance is applied separately to each pair of matched columns. Thus $\bM_J$ is unique up to signed permutations of the columns. Next, $\bM_{Ix}$ and $\bM_{Iy}$ correspond to the columns having chordal distance greater than zero for all possible matchings. The columns of  $\bM_{Nx}$ and $\bM_{Ny}$ are not unique. 

To summarize, the probabilistic model for joint structure posits the existence of a linear transformation $\bM_J$ (i.e., mixing) that acts on both $\bs_{Jx}$ and $\bs_{Jy}$, whereas the individual structure is subject to different linear transformations. 

Compared to the measurement error model where $\bn_{x} \in \tR^n$, here $\bn_{x} \in \tR^{n-r_x-1}$. 
The rank of the noise influences 1) whether the non-Gaussian signals can be recovered without noise, and 2) whether the PCA step in ICA methods discards non-Gaussian components. To see this, consider the LNGCA and measurement error models for a single dataset differing only in the Gaussian components: $\bx = \bM_x \bs + \bM_n \bn$ and $\bx^* = \bM_x \bs + \bM_n^* \bn^*$, where the latter is the measurement error model with full rank $\bM_n^* \in \tR^{n \times n}$. Then define $\bM_x^-$ such that $\bM_x^- \bM_x = \bI_{r_x}$. In LNGCA, we define $\bM_x^- = \bM_x^\top \bL^2$ where $\bL$ is the square root of the generalized inverse of $\mathrm{Cov}(\bx)$. Then it can be shown that $\bM_x^- \bM_n = \bzero$. In contrast, $\bM_n^*$ spans $\tR^n$ in the measurement error model, so $\bM_x^- \bM_n^*  \ne \bzero$ for all $\bM_x^-$. Consequently, $\bs$ is corrupted by noise even if the true $\bM_x$ and $\bM_n^*$ were known. 
Second, the eigenvalues of $\mathrm{Cov}(\bx)$ from  \eqref{eq:probmodel} capture the largest directions of variance, which are in general not related to the non-Gaussian signal. Thus a probabilistic PCA decomposition used in ICA \citep{beckmann2004probabilistic} can fail to extract the non-Gaussian subspace. If one instead assumes full rank Gaussian noise, the probabilistic SING model in \eqref{eq:probmodel} can provide a useful approximation; this is related to the noisy ICA model approximated by LNGCA. In fact, LNGCA can also be more effective than PCA+ICA for full rank noise (Figure S.2 in \cite{risk2017linear}). Also note in the measurement error model, higher variance directions of noise can dominate the PCA if the noise is not isotropic, and LNGCA may be more effective. 
\vspace{0.5cm}

\emph{Remark 2.} In LNGCA, maximizing the JB statistic provides a consistent estimator of the LNGCA model under finite eighth moments of the components $\bs_j$ for an independent and identically distributed sample $j=1,\dots,p$; see Theorem 5.2 in \cite{virta2016projection}. Let $\bx_j \in \tR^n$, $j=1,\dots,p_x$, be an iid sample from \eqref{eq:probmodel}. Then we obtain a consistent estimate of the object $\bM_x = [\bM_{J} \bD_x, \bM_{Ix}]$ up to signed permutations of the columns as $p_x \rightarrow \infty$ for fixed $n$. Similarly for $\by_k \in \tR_n$, $k=1,\dots, p_y$. Since the estimators of $\bM_x$ and $\bM_y$ are consistent, the sum of the chordal distances between matched columns of $\bM_x$ and $\bM_y$ converges to zero. Hence, we obtain a consistent estimator of $\bM_J$. This motivates the initial estimator used in Section \ref{sec:2.2} and described in Appendix \ref{appendix:InitialAlgorithm}. We acknowledge that this is non-standard from a probabilistic perspective, but reiterate that maximizing non-Gaussianity across features has been used in thousands of neuroimaging studies \citep{calhoun2012multisubject}, and this remark provides a perspective on this approach in the context of \eqref{eq:model} and related integration methods like Joint ICA.
\vspace{0.5cm}

\subsection{Simultaneous NGCA fitting and algorithm}\label{sec:2.2}
Recall the whitening matrix for $\bX_c$ is $\hatbL_x$ and define $\hatbL_x^{-1} = (\bX_c \bX_c^{\top}/n)^{1/2} = \bV_x \bLambda_x^{1/2} \bV_x^\top$. Let $\bX_w = \hatbL_x \bX_c$. Similarly define the whitening matrix $\widehat \bL_y$ for $\bY_c$ and whitened data $\bY_w$. Note that for an estimate $\widehat \bM_x$, we can write $\widehat \bM_x = \widehat \bL_x^{-1}\widehat \bU_x^{\top}$, where $\bU_x^\top$ is semi-orthogonal. Similarly, $\widehat \bM_y = \widehat \bL_y^{-1}\widehat \bU_y^{\top}$. Let $f$ be the JB statistic as defined in~\eqref{eq:JB}. We consider
\begin{equation}\label{eq:joint_app3}
\begin{split}
    \minimize_{\bU_x, \bU_y}&\Big[-\sum_{l=1}^{r_x}f(\bu_{xl}^{\top}\bX_w)-\sum_{l=1}^{r_y} f(\bu_{yl}^{\top}\bY_w)+\rho\sum_{l=1}^{r_J} d(\widehat \bL_x^{-1}\bu_{xl}, \widehat \bL_y^{-1}\bu_{yl})\Big\}\Big]\\
    \mbox{subject to}&\quad \bU_x\bU_x^{\top} = \bI_{r_x}, \quad \bU_y\bU_y^{\top} = \bI_{r_y},
    \end{split}
\end{equation}
where $d(\bx,\by)$ is the chosen distance metric between vectors $\bx$ and $\by$. When $\rho=0$, problem~\eqref{eq:joint_app3} reduces to two separate LNGCA optimization problems \eqref{eq:LNGCAsep} with $r_x$ and $r_y$ components, respectively. When $\rho \to \infty$, the constraint leads to equality of the first $r_J$ columns in $\widehat \bM_x$ and $\widehat \bM_y$ with respect to the chosen distance metric, thus the first $r_J$ columns, rescaled to have norms equal to one, provide an estimate of $\widehat \bM_J$. In practice, we recommend choosing $\rho$ to result in columns that are approximately equal. In our data application, this is satisfied by setting $\rho$ equal to the sum of the JB statistics of all joint components from the separate LNGCA divided by 10; see Section \ref{sec:hcpdata}. Because model~\eqref{eq:model} allows for different scales between datasets, we choose a distance metric that is scale and sign invariant - the squared chordal distance~\eqref{eq:distance}.
 Since each column of $\bX_c$ and $\bY_c$ has mean zero, a chordal distance equal to zero corresponds to correlation equal to one. 

There are two difficulties in solving problem~\eqref{eq:joint_app3}: it has a nonconvex objective function and it has two orthogonality constraints. In ICA, it is common to use a fixed point algorithm where each iteration includes an approximate Newton update to the unmixing matrix $\bU$, followed by an orthogonalization step to project the unmixing matrix back to the Stiefel manifold \citep{Hyvarinen:1999ek}. However, the algorithm can not be easily generalized to our case due to the addition of a distance metric on the mixing matrices. Instead, we propose to modify the curvilinear search algorithm of \citet{Wen:2012ga}. The algorithm is designed for problems with differentiable objective functions and orthogonality constraints. The main advantage of the algorithm is that it performs feasible updates, unlike the fixed-point algorithm, and therefore does not need an extra projection step onto the manifold. 

A direct application of the algorithm to problem~\eqref{eq:joint_app3} is to consider alternating minimization over $\bU_x$ and $\bU_y$. While this works well for moderately-sized problems, we found that it is possible to modify the algorithm to perform joint updates over $\bU_x$ and $\bU_y$, which leads to significant speed improvements. Let $G_x(\bU_x, \bU_y)$ be the gradient of the objective function with respect to $\bU_x$ evaluated at the current values of $\bU_x$ and $\bU_y$, and similarly define $G_y(\bU_x, \bU_y)$. The value of $\rho$ affects both $G_x(\cdot)$ and $G_y(\cdot)$. The proposed optimization algorithm for~\eqref{eq:joint_app3} is summarized in Algorithm~\ref{a:algorithm}, and the full derivation together with the discussion of alternatives and step size selection is presented in the Appendix~\ref{appendix:algorithm1}. To monitor convergence, we use $\sqrt{\textrm{PMSE}}$ as defined in~\eqref{eq:pmse}.

Let $\widehat \bU_x$ and $\widehat \bU_y$ be the values of $\bU_x$ and $\bU_y$ at convergence. The corresponding estimated non-Gaussian components are defined as $\widehat \bS_x = \widehat \bU_x \bX_w$, $\widehat \bS_y = \widehat \bU_y \bY_w$. Then the first $r_J$ columns of $\hatbM_x = \hatbL_x^{-1} \widehat \bU_x^\top$ scaled to unit norm provide an estimate of $\hatbM_{J}$, and for sufficiently large $\rho$, this is equal up to scaling to the first $r_J$ columns of $\hatbM_Y$. Additionally, the first $r_J$ rows of $\hatbS_x$ correspond to $\hatbS_{Jx}$.

\begin{algorithm}[!t]
\DontPrintSemicolon

\KwInput{$\bX_w$, $\bY_w$, $\widehat \bL_x^{-1}$, $\widehat \bL_y^{-1}$, $\bU_x^{(0)}$, $\bU_y^{(0)}$, $\rho>0$, $\varepsilon > 0$}

$G_x(\bU_x, \bU_y), G_y(\bU_x, \bU_y)$ - gradients of objective function with respect to $\bU_x$ and $\bU_y$, respectively, at current values of $\bU_x$, $\bU_y$; depend on the \textbf{Input}

$k = 0$;

\Repeat{$\sqrt{\text{PMSE}(\bU_x^{(k-1)},\bU_x^{(k)})} + \sqrt{\text{PMSE}(\bU_y^{(k-1)},\bU_y^{(k)})}<\varepsilon$}{
    $k = k+1$;
     
    $\bW_x = {\bU_x^{(k-1)}}^{\top}G_x(\bU_x^{(k-1)}, \bU_y^{(k-1)})^{\top} - G_x(\bU_x^{(k-1)}, \bU_y^{(k-1)})\bU_x^{(k-1)}$;
    
    $\bW_y = {\bU_y^{(k-1)}}^{\top}G_y(\bU_x^{(k-1)}, \bU_y^{(k-1)})^{\top} - G_y(\bU_x^{(k-1)}, \bU_y^{(k-1)})\bU_y^{(k-1)}$;
    
    Select step size $\tau_k >0$;
   
    $\bU_x^{(k)} = \bU_x^{(k-1)}\Big(\bI-\frac{\tau_k}2\bW_x\Big)\Big(\bI+\frac{\tau_k}2\bW_x\Big)^{-1}$;
    
    $\bU_y^{(k)} = \bU_y^{(k-1)}\Big(\bI-\frac{\tau_k}2\bW_y\Big)\Big(\bI+\frac{\tau_k}2\bW_y\Big)^{-1}$;
}
\caption{Curvilinear search algorithm for~\eqref{eq:joint_app3}, see Appendix~\ref{a:algorithm} for additional details.}\label{a:algorithm}
\end{algorithm}

Since problem~\eqref{eq:joint_app3} is nonconvex, Algorithm~\ref{a:algorithm} requires careful initialization of $\bU_x^{(0)}$ and $\bU_y^{(0)}$. We initialize the algorithm with the solutions obtained by solving two separate NGCA problems (problem~\eqref{eq:joint_app3} with $\rho = 0$) using a fixed-point algorithm with multiple random starting points. Then the columns in $\bU_x^{(0)}$ are reordered to match the columns in $\bU_y^{(0)}$ based on greedy pairwise matching of chordal distances between corresponding $\widehat \bM_x^{(0)}$ and $\widehat \bM_y^{(0)}$, as described in Appendix~\ref{appendix:InitialAlgorithm}. 
A permutation test is used to determine whether the correlation between matched columns is significant, as described in Section~\ref{sec:rJ}. 

\subsection{Selecting the number of components}\label{sec:selection}

\subsubsection{Signal rank: choosing \texorpdfstring{$r_x$}{rx}}

Resampling and asymptotic tests for the signal rank of the non-Gaussian subspace in LNGCA have been developed \citep{nordhausen2016asymptotic,zinlngca}. We also propose a permutation test that has some computational advantages as each iteration estimates a single non-Gaussian component  (Appendix~\ref{appendix:rank}). We can further reduce computational expense by estimating the ranks of the joint components only (next section) using as input the estimate of the saturated LNGCA model, i.e., where the number of components is equal to the rank of the data. This approach works well for recovering the joint components, which is our focus here. 

\subsubsection{Joint rank: choosing \texorpdfstring{$r_J$}{rJ}}\label{sec:rJ}
Let $\hatbM_x^{u}$ and $\hatbM_y^{u}$ be the estimates of the mixing matrices from the separate LNGCA decompositions of $\bX$ and $\bY$, which contain an unordered concatenation of estimates of joint and individual subject scores. We use a greedy algorithm that sequentially matches a column of $\hatbM_x^{u}$ with $\hatbM_y^{u}$ by minimizing chordal distance, removes the matched columns, and then finds the next pair. This is equivalent to maximizing absolute correlation for data in which each column has mean equal to zero. Let $\hatbM_x^{(0)}$ and $\hatbM_y^{(0)}$ denote the ordered mixing matrices. Next, we determine which pairs of columns are sufficiently close to be deemed estimates of joint subject scores. Let $\psi_r$ be the chordal distance between the matched columns for the $r$th pair of components, $r=1,\dots,n-1$. For permutations $t=1,\dots,T$, we fix $\hatbM_x^{(0)}$ and generate $\hatbM_y^{[t]} = \bP^{[t]} \hatbM_y^{(0)}$, where $\bP^{[t]}$ is a random permutation matrix. Let $\psi_{lm}^{[t]}$ be the distance between the $l$th column of $\hatbM_x^{(0)}$ and the $m$th column of $\hatbM_y^{[t]}$. Define $\psi_{min}^{[t]} = \min_{l=\{1,\dots,r_x\},m=\{1,\dots,r_y\}} \psi_{lm}^{[t]}$. Then an FWER-adjusted p-value for the $r$th component is $\frac{1}{T} \sum_{t} \iv (\psi_r > \psi_{min}^{[t] })$, where $\iv$ is the indicator function. We set $r_J$ equal to the largest index with $p<\alpha$, e.g., $\alpha=0.01$. For a given joint component, this approach assumes the chordal distance between two correctly matched columns will be less than the distance between mismatched columns.

\section{Simulations}\label{sec:Simulations}

We consider the following methods for comparison: Joint ICA, mCCA+jICA \citep{sui2011discriminating}, separate LNGCA (corresponding to $\rho = 0$), and SING with three values of $\rho$ designed to examine the impact of this tuning parameter on accuracy. In addition, we consider an alternative approach to fitting the SING model based on averaging the mixing matrices from separate LNGCA models (SING-averaged). Given $\hatbM_J$, this approach estimates the non-Gaussian components using the Procrustes solution to result in orthogonal components, in contrast to least-squares. The detailed description is in the Appendix~\ref{sec:averageSING}.
We use our own implementation of Joint ICA and mCCA+jICA. To improve comparability, we use the same measure of non-Gaussianity (JB statistic) in all approaches. All analyses were implemented in R. 

We evaluate the quality of estimation of non-Gaussian components and mixing matrices based on scale and permutation invariant mean squared error set forth in \citet{risk2017linear}. Let $\bS\in \R^{p\times r}$ be the matrix of true non-Gaussian components, and let $\widehat \bS\in \R^{p \times r}$ be the corresponding estimate. We assume that both $\bS$ and $\widehat \bS$ are scaled to have column variance one. Let $\mathcal{P}_{\pm}$ be the class of $r \times r$ signed permutation matrices. The permutation-invariant mean squared error is defined on scaled $\bS$ and $\widehat \bS$ as
\begin{equation}\label{eq:pmse}
\mathrm{PMSE}(\bS, \widehat \bS) = \frac1{rp}\argmin_{\bP\in \mathcal{P}_{\pm}}\|\bS - \bP\widehat \bS\|_F^2,  
\end{equation}
and we report error as $\sqrt{\mbox{PMSE}}$. The error for mixing matrices $\bM$ and $\widehat \bM$ is calculated similarly, where scale invariance is achieved using column scaling.

In addition, we evaluate the overall error in estimating joint structure in model~\eqref{eq:model}. Define $\bJ_x = \bM_J\bD_x\bS_{J x}$, and similarly $\bJ_y = \bM_J\bD_y\bS_{J y}$. Then the corresponding estimates are $\widehat \bJ_x = \widehat \bM_x\widehat \bS_x$, and similarly $\widehat \bJ_y$. The mean squared error is defined as
\begin{equation}\label{eq:mse}
\mathrm{MSE}(\bJ, \widehat \bJ) = \frac{\|\bJ - \widehat \bJ\|_F^2}{\|\bJ\|_F^2},  
\end{equation}
and we report this error as $\sqrt{\mbox{MSE}}$.

An important factor impacting the accuracy of estimates of the mixing matrix and/or non-Gaussian components is the amount of variance in the non-Gaussian subspace. We decompose the total variance into joint, individual, and noise contributions. Let $\|\bX\|_F$ be the Frobenius norm of $\bX$. Then define
\begin{align}\label{eq:R2}
R_{Jx}^2 = \|\bM_J \bD_x \bS_{Jx}\|_F^2 / \|\bX\|_F^2 = \|\frac{1}{p_x} \bX \bS_{Jx}^\top \bS_{Jx}\|_F^2/\|\bX\|_F^2,
\end{align}
and $R_{Ix}^2 = \|\frac{1}{p_x}\bX \bS_{Ix}^\top \bS_{Ix}\|_F^2/\|\bX\|_F^2$. With the orthogonality conditions in \eqref{eq:model}, the total non-Gaussian signal variance is $R_{Sx}^2 =R_{Jx}^2+R_{Ix}^2$. We also define the signal variance to noise variance ratio (SNR) for each dataset as the ratio of the variance in the non-Gaussian subspace to the variance in the Gaussian subspace. Letting $R_{Nx}^2$ denote the proportion of variance in the Gaussian subspace, $\text{SNR} = R_{Sx}^2 / R_{Nx}^2$, which is equivalent to the SNR definition in \cite{risk2017linear}.

\subsection{Simulation Setting 1}\label{sec:sim_small}
\begin{figure}[!t]
    \centering
    \includegraphics[width =0.6\textwidth]{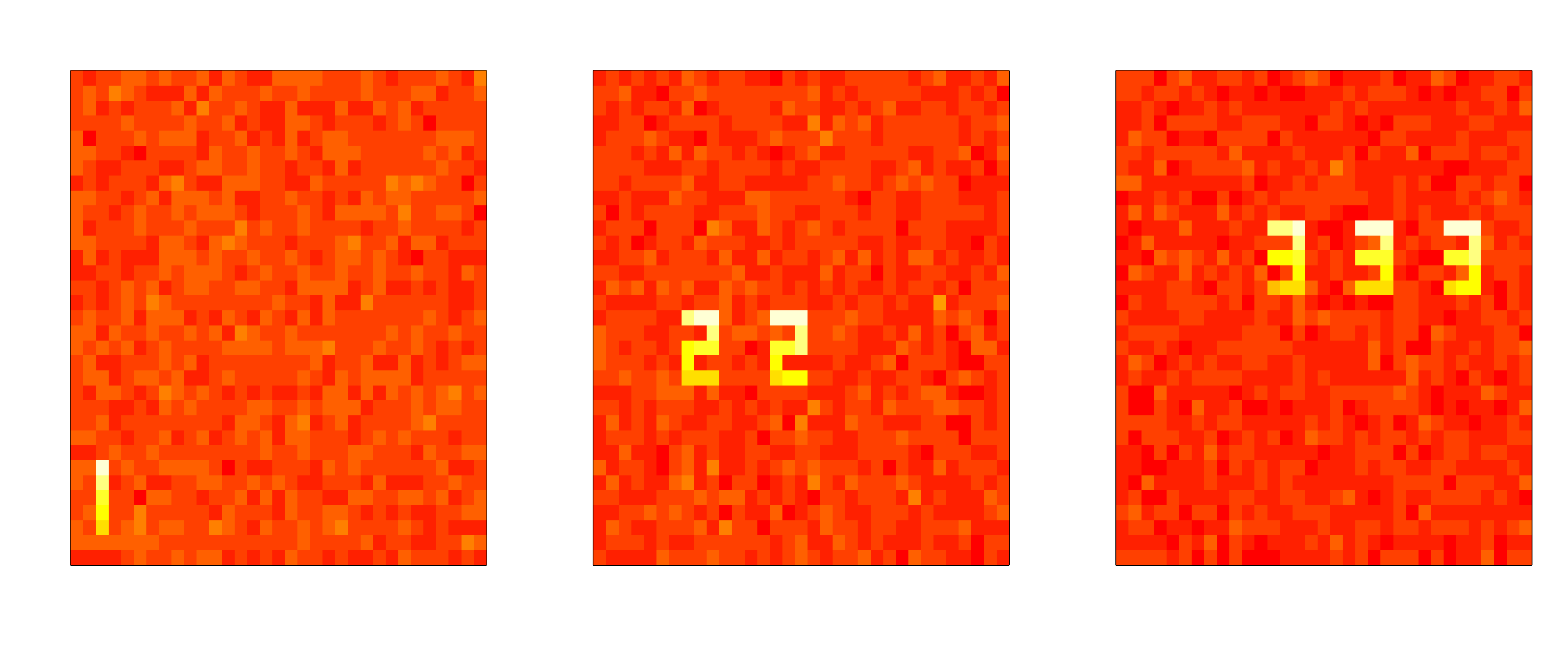}
    \includegraphics[width =0.6\textwidth]{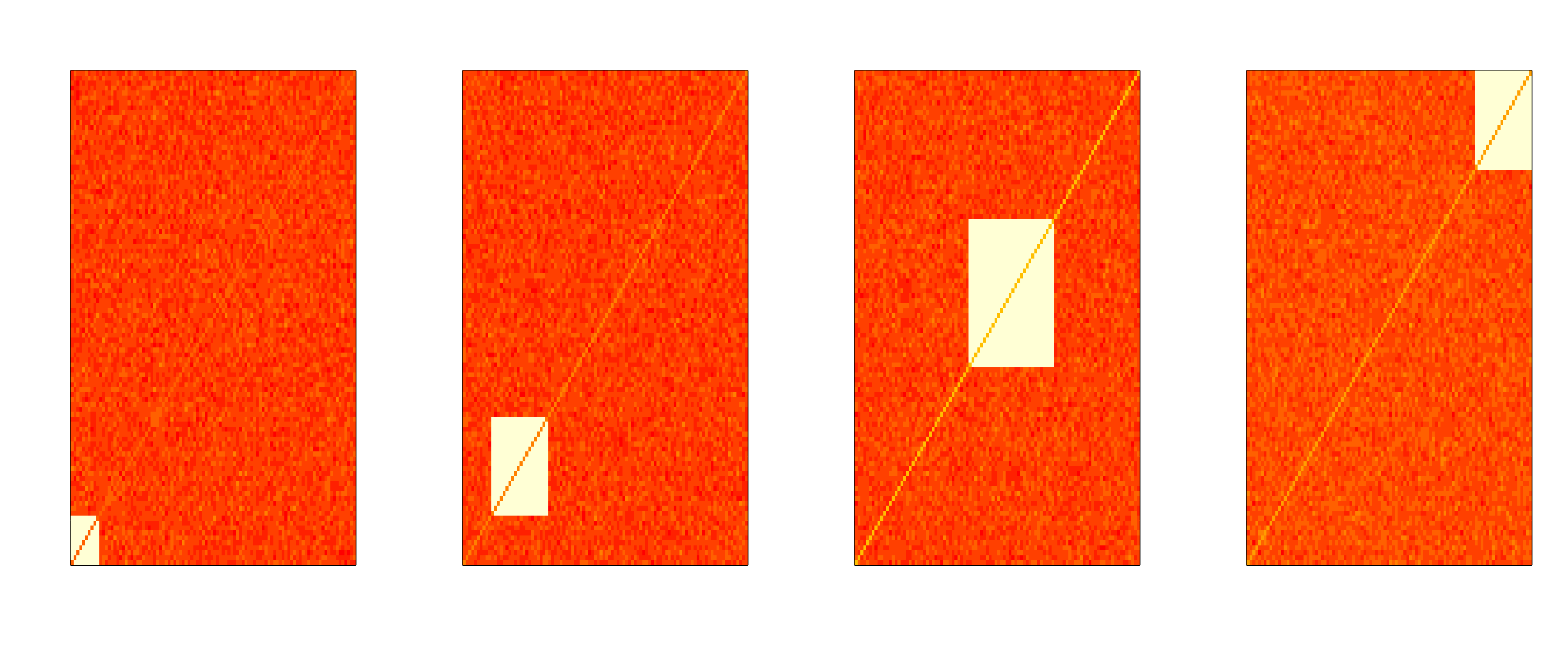}
    \caption{Top: True components for $\bX$, first is individual, and the remaining two are shared. Bottom: True components for $\bY$, first two are shared, and the remaining two are individual.}
    \label{fig:components_simul}
\end{figure}

In this section, we evaluate the estimation accuracy of SING when data are generated from \eqref{eq:model} with the number of subjects $n=48$. We use 3 non-Gaussian components for dataset $\bX$ and 4 non-Gaussian components for dataset $\bY$, with $r_J=2$. Figure~\ref{fig:components_simul} displays all the components for each dataset. The components $\bS_x$ are vectorized $33 \times 33$ images, corresponding to 1089 features. The components $\bS_y$ are the lower-diagonals of $100 \times 100$ symmetric matrices, corresponding to 4950 features, in which a contiguous block has values equal to 1, and the other values are generated from a mean zero normal distribution with variance $\sigma^2 = 0.005$. The spatially adjacent features in $\bS_x$ and the block features in $\bS_y$ share similarities with the Joint ICA decomposition of task maps and resting-state correlations in \citealt{lerman2017multimodal}.

For each simulation, we generate the joint subject scores $\bM_J = [\bm_{J1}, \bm_{J2}]\in \R^{n \times 2}$ with  $\bm_{J1} \sim N(\bmu_1, \bI_n)$, $\bm_{J2} \sim N(\bmu_2, \bI_n)$, $\bmu_1 = (\ones_{24}^\top, - \ones_{24}^\top)^{\top}$ and $\bmu_2 = (-\ones_{24}^\top, \ones_{24}^\top)^{\top}$. 
We set $\bD_x = \bI$ and $\bD_y = \mbox{diag}(-5,2)$ to have difference in both sign and scale between the two datasets. We generate $\bM_{Ix}$ and $\bM_{Iy}$ similar to $\bM_J$ using iid unit variance Gaussian entries with means equal to $\bmu_{3x} =  (-\ones_{12}^{\top},\ones_{12}^{\top},-\ones_{12}^{\top}, \ones_{12}^{\top})^{\top}$, $\bmu_{3y} = (-\ones_6^{\top},\ones_{6}^{\top},-\ones_{6}^{\top}, \ones_{6}^{\top}, -\ones_6^{\top},\ones_{6}^{\top},-\ones_{6}^{\top}, \ones_{6}^{\top})^{\top}$, $\bmu_{4y} = (\ones_{24}^{\top}, -\ones_{24}^{\top})^{\top}$. These means result in various degrees of correlation between the columns of the mixing matrices. For the Gaussian noise, we generate $\bM_{Nx}$, $\bM_{Ny}$, $\bN_x$ and $\bN_y$ using iid Gaussian mean zero entries. We set the noise variance to reach the pre-specified SNR regime as defined in Section \ref{sec:model}. We consider a crossed design for the SNR in $\bX$ and $\bY$ with low SNR = 0.2 and high SNR = 5, specifically 1) low SNR $\bX$, low SNR $\bY$; 2) low SNR $\bX$, high SNR $\bY$; 3) high SNR $\bX$ and low SNR $\bY$; and 4) high SNR $\bY$ and high SNR $\bX$. This results in $R^2_{Jx} \in [0.09, 0.13]$ (low SNR $\bX$), $R^2_{Jx} \in [0.40, 0.65]$ (high SNR $\bX$), $R^2_{Jy} \in [0.15, 0.16]$ (low SNR $\bY$), and $R^2_{Jy} \in [0.72, 0.80]$ (high SNR $\bY$). 

We first evaluate the performance of the proposed scheme for selecting the number of joint components (Section~\ref{sec:rJ}) over 100 simulations for each SNR combination. As described in Section \ref{sec:rJ}, we estimated the saturated model for each dataset using 20 restarts and $\alpha = 0.01$. Here we did not perform double centering since the true scores have sample mean approximately equal to zero, and hence the saturated model corresponds to $r_x=r_y=48$. We found that this leads to correct $\widehat r_J=2$ in 397 out of 400 simulations. In the remaining 3 simulations (1 for low SNR $\bX$, high SNR $\bY$; 2 for high SNR $\bX$, high SNR $\bY$), the $r_J$ is slightly over-estimated with $\widehat r_J =3$.

\begin{figure}[!t]
    \centering
    \includegraphics[width=0.75\textwidth]{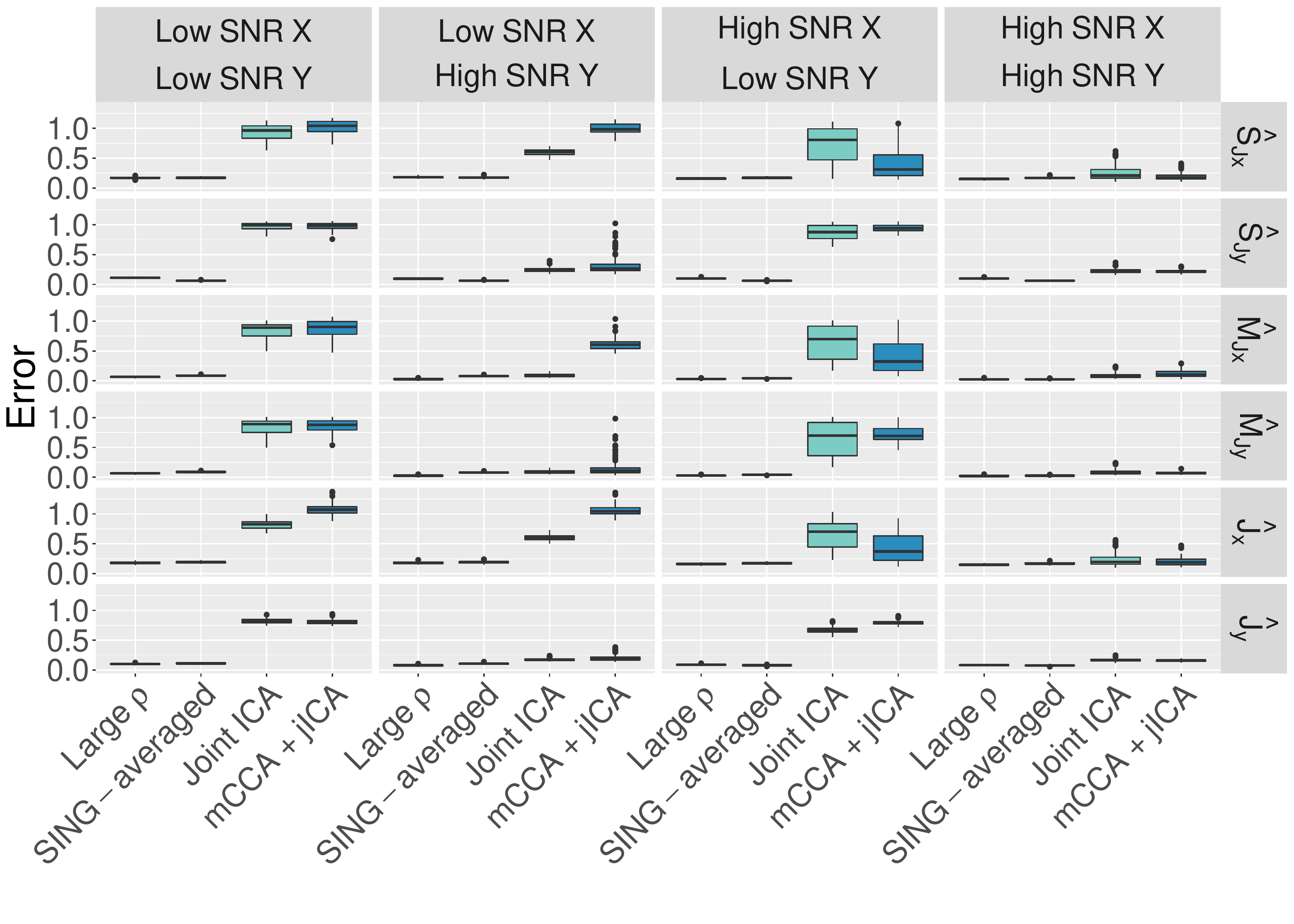}\\
    \includegraphics[width=0.75\textwidth]{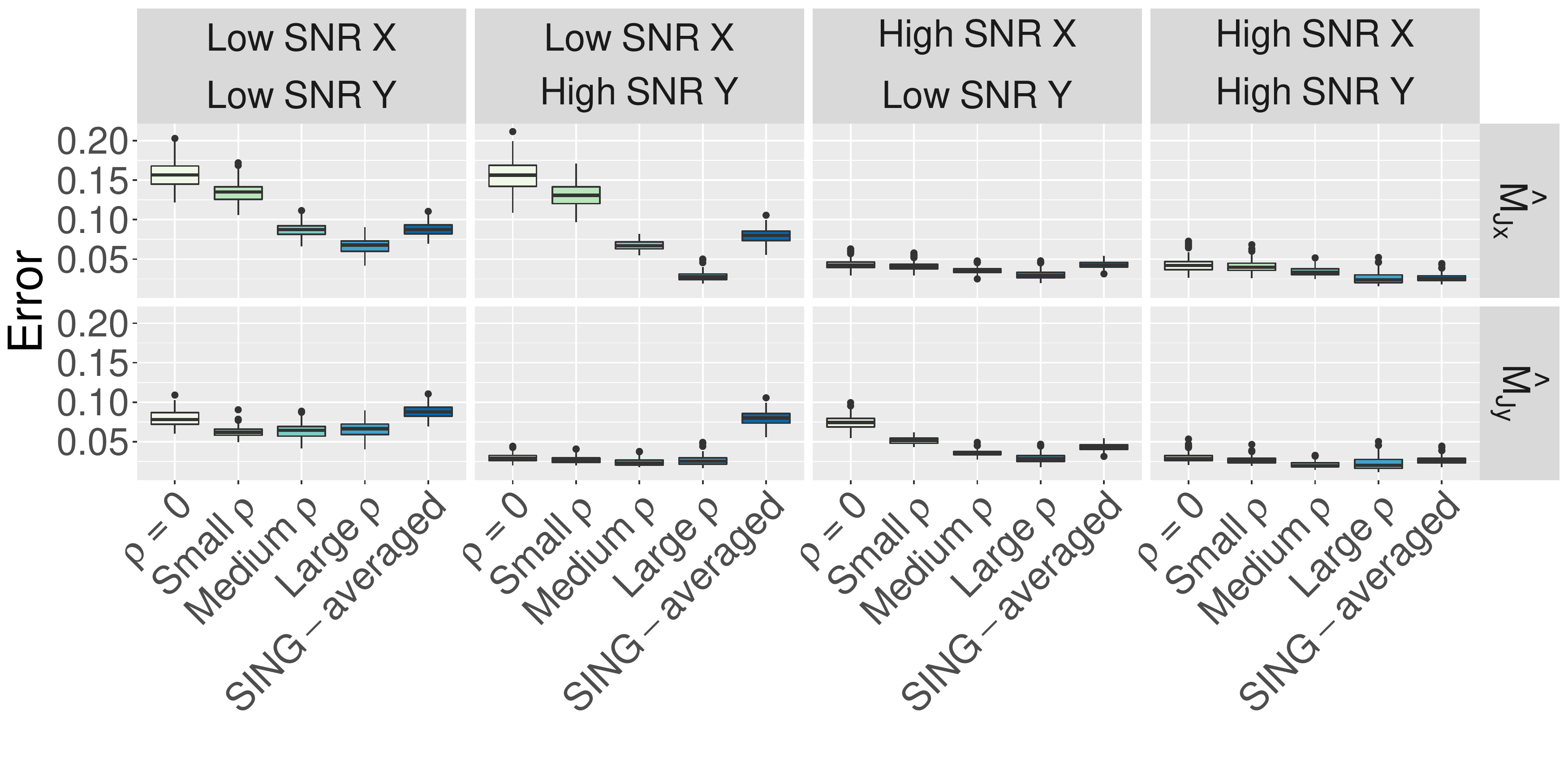}
    \caption{Simulation Setting 1 of Section~\ref{sec:sim_small}, results are over 100 replications for each combination of signal to noise ratio (SNR). Top: results from SING with large $\rho$, SING-averaged from $\rho=0$, Joint ICA, and mCCA+jICA. The errors are evaluated using $\sqrt{PMSE}$ from~\eqref{eq:pmse}  (components and mixing matrices). By definition, $\sqrt{\text{PMSE}}\in [0, \sqrt{2}] = [0, 1.414]$. The errors on $\widehat \bJ_x$ and $\widehat \bJ_y$ are evaluated using $\sqrt{\text{MSE}}$ from~\eqref{eq:mse}. Bottom: the impact of $\rho$ on subject score estimation in SING variants.}
    \label{fig:sim_small}
\end{figure}

We next consider the quality of estimation of non-Gaussian components and mixing matrices. Let $\hat{\rho}$ equal the sum of the JB statistics of joint components from the separate analyses. We compare multiple estimation schemes: 1)~Joint ICA with the true number of joint components $r_J=2$, where we follow the default normalization in \cite{rachakonda2012fusion} in which each dataset is scaled by the square root of the mean of squared elements before applying Joint ICA; 2)~mCCA+jICA with true $r_x=3$, $r_y=4$ and $r_J=2$, where each dataset is scaled by the square root of mean of squared elements as in Joint ICA; 3)~separate analysis ($\rho=0$) with true $r_x=3$ and $r_y=4$, and subsequent matching as in Section~\ref{sec:rJ} to determine $2$ joint components, 4)~small~$\rho$, set equal to $0.1\hat{\rho}$, 5)~medium~$\rho = \hat{\rho}$, 6)~large~$\rho = 20\hat{\rho}$,  and 7)~SING-averaged from $\rho=0$ (Appendix~\ref{sec:averageSING}). For Joint ICA, mCCA+jICA, and the separate analysis with $\rho = 0$, we use 20 random starting points to choose the solution that optimizes the objective function. For $\rho>0$, we use the solution obtained at $\rho = 0$ as a starting point. Figure~\ref{fig:sim_small} shows the results for all methods. Although in truth $\bM_{Jx}=\bM_{Jy}=\bM_J$, we report separate estimates of $\bM_J$ corresponding to the estimate from $\bX_c$ and $\bY_c$, denoted $\hatbM_{Jx}$ and $\hatbM_{Jy}$, since these estimates are not in general equal in SING or mCCA+jICA. 

All variants of SING perform reasonably well across SNR settings with improvements with larger $\rho$, while Joint ICA and mCCA+jICA are effective only in the high SNR $\bX$ and high SNR $\bY$ setting  (Figure \ref{fig:sim_small}, top panel). For both Joint ICA and mCCA+jICA, the methods perform poorly for low SNR because the Gaussian noise components dominate the non-Gaussian components (top panel, column 1). Joint ICA uses PCA after concatenating standardized datasets, and mCCA applies PCA separately to each dataset prior to CCA on the subject scores. In the low SNR $\bX$ and low SNR $\bY$ scenarios, the non-Gaussian directions are discarded. In the high SNR $\bY$ and low SNR $\bX$ setting  (top panel, column 2), the performance of Joint ICA improves particularly for $\hatbS_{Jy}$, $\hatbM_{Jx}$, $\hatbM_{Jy}$, $\hatbJ_x$, and $\hatbJ_y$.  This is likely due to the fact that $\bX$ has only 1089 features while $\bY$ has 4950 features, and thus signal in $\bY$ dominates the SVD of the concatenated data that is at the core of Joint~ICA. However, mCCA+jICA is only accurate with respect to $\hatbS_{Jy}$, $\hatbM_{Jy}$, and $\hatbJ_y$. In the high SNR $\bX$ and low SNR $\bY$ setting (column 3), in Joint ICA, the noisy $\bY$ dominates the SVD, and the overall performance is poor across all $\hatbS_{Jx}$, $\hatbS_{Jy}$, $\hatbM_{Jx}$, $\hatbM_{Jy}$, $\hatbJ_x$, and $\hatbJ_y$. Similar to the previous setting, mCCA+jICA performs better with respect to the high SNR dataset, namely $\hatbS_{Jx}$, $\hatbM_{Jx}$, and $\hatbJ_x$, but poorly for $\hatbS_{Jy}$, $\hatbM_{Jy}$, and $\hatbJ_y$. In the high SNR $\bX$ and high SNR $\bY$ scenarios, both Joint ICA and mCCA+jICA perform well across $\hatbS_{Jx}$, $\hatbS_{Jy}$, $\hatbM_{Jx}$, $\hatbM_{Jy}$, $\hatbJ_x$, and $\hatbJ_y$. This is because the non-Gaussian signal is now contained in the directions of large variance, and hence the PCA steps are appropriate.

In the bottom panel in Figure \ref{fig:sim_small}, we see that increasing $\rho$ leads to improvements with respect to $\hatbM_{Jx}$ and $\hatbM_{Jy}$, as the SING algorithm leads to effective data integration. The estimates of $\hatbM_J$ improve uniformly as $\rho$ increases, where large values of $\rho$ result in $\hatbM_{Jx} \approx \hatbM_{Jy}$. In particular, in the low SNR $\bX$ and high SNR $\bY$ setting, we see improvements in $\hatbM_{Jx}$ in addition to $\hatbM_{Jy}$. In the high SNR $\bX$ and low SNR $\bY$, we see improvements in $\hatbM_{Jy}$ in addition to $\hatbM_{Jx}$. Relative to SING with $\rho>0$, SING-averaged is competitive but performs worse than large $\rho$, particularly for low SNR $\bX$ and high SNR $\bY$. In this setting, comparing $\hatbM_{Jx}$ and $\hatbM_{Jy}$ for SING with $\rho=0$, we see greater accuracy for $\hatbM_{Jy}$ than $\hatbM_{Jx}$. Then as we increase $\rho$, SING appears to leverage the greater accuracy in $\hatbM_{Jy}$ to improve $\hatbM_{Jx}$. In contrast, SING-average treats the two datasets as having equal information, and so the noisier $\hatbM_{Jx}$ leads to overall reduced accuracy of the average. In Figure~\ref{fig:SINGsetting1}, we see that as $\rho$ increases, there is a small cost in terms of the accuracy of $\hatbS_{Jy}$, although all SING variants are accurate. The small decline in accuracy for large $\rho$ may be related to the fact that the penalized objective can lead to components that have lower non-Gaussianity compared to the unpenalized formulation. 

In summary, relative to the low SNR setting, Joint ICA can improve subject scores in the mixed SNR setting when the larger dataset (with respect to the number of features) has high SNR; mCCA can improve subject scores in the mixed SNR setting for the dataset containing high SNR; and SING can improve subject scores across SNR settings.

\subsection{Simulation Setting 2}\label{sec:sim_large}

We consider a large-scale simulation setting with $n=48$ subjects, $p_x = 59,412$, and $p_y = 71,631$. We generate a dataset from \eqref{eq:model} with 12 non-Gaussian components for each $\bX$ and $\bY$. To mimic the non-Gaussian components from real data, we select components estimated from the HCP data as the truth. We fix $r_J=2$ shared subject score directions, and $\bM_J\bD_x$ and $\bM_J\bD_y$ are generated as in Section~\ref{sec:sim_small}. The elements of  $\bM_{Ix}$ and $\bM_{Iy}$ are generated using independent normals with zero mean and standard deviations 10 and 30, respectively. For the Gaussian noise, we generate $\bM_{Nx}$, $\bM_{Ny}$, $\bN_x$ and $\bN_y$ using iid Gaussian mean zero entries, which are then rescaled to reach SNR$_{x}$ = SNR$_{y}$ = 0.5. The resulting proportions of joint variance explained are $R^2_{Jx} = 0.0014$ and $R^2_{Jy} = 0.0021$, which is similar to the values observed in the HCP data (Section~\ref{sec:data}).
First, we investigate the performance of the joint rank selection procedure from Section~\ref{sec:rJ}. We consider two scenarios: (i) estimation based on the saturated model; (ii) estimation based on the model with the true number of components with $r_x = r_y = 12$. In both cases, the true $r_J = 2$ is selected for $\alpha \in \{0.01, 0.05, 0.1\}$.

Next, we compare the estimation performance between Joint ICA, mCCA+jICA, and SING variants. As before, let $\hat{\rho}$ equal the sum of the JB statistics of joint components from the separate analyses. We consider 1) separate analysis with matched components ($\rho=0$); 2) small $\rho = \hat{\rho}/10$; 3) medium $\rho = \hat{\rho}$; 4) large $\rho = 10 \hat{\rho}$; and 5) SING-averaged from $\rho = 0$ as described in the Appendix~\ref{sec:averageSING}.
For the algorithm initialization at $\rho>0$, we input the first 12 components estimated from the separate analyses and set $\hat{r}_J=2$. Table~\ref{tab:sim_large} displays the errors of $\hatbS_{Jx}$, $\hatbS_{Jy}$, $\hatbM_{Jx}$, $\hatbM_{Jy}$, $\hatbJ_x$, and $\hatbJ_y$ for Joint ICA, mCCA+jICA, and the SING variants. The conclusions are similar to the low SNR results in ~Section~\ref{sec:sim_small}. Both Joint ICA and mCCA+jICA perform poorly compared to the SING variants.  In this setting, the small, medium, and large $\rho$ all lead to $\hatbM_{Jx} \approx \hatbM_{Jy}$, and the accuracies are similar. Compared to $\rho=0$, $\hatbM_{Jx}$ and $\hatbM_{Jy}$ are more accurate with $\rho>0$. Compared to SING-averaged, $\hatbS_{Jx}$ and $\hatbS_{Jy}$ are more accurate with $\rho>0$. The results in this simulation setting are insensitive to the sizes of $\rho>0$ evaluated here, where all the choices resulted in $\hatbM_{Jx} \approx \hatbM_{Jy}$. 

\begin{figure}[!t]
    \centering
    \includegraphics[width=\textwidth]{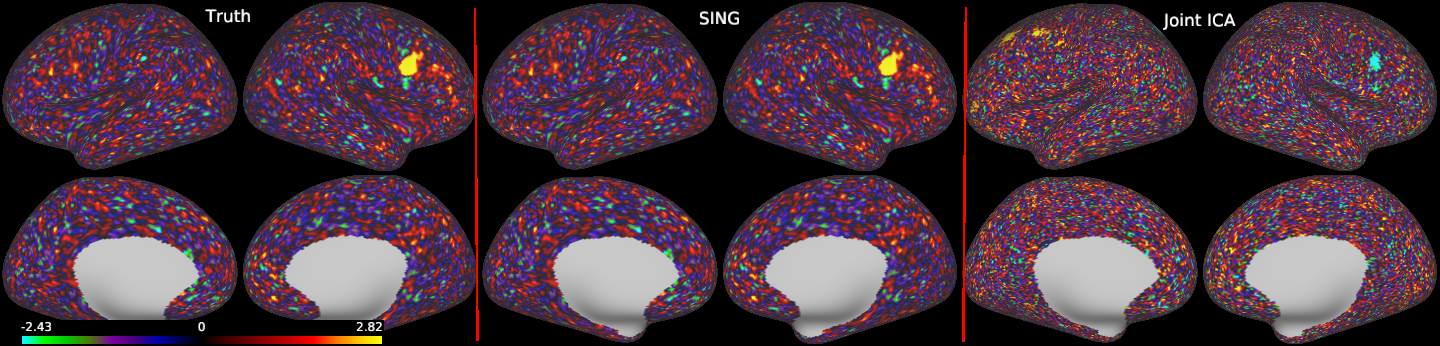}
    \includegraphics[width=\textwidth]{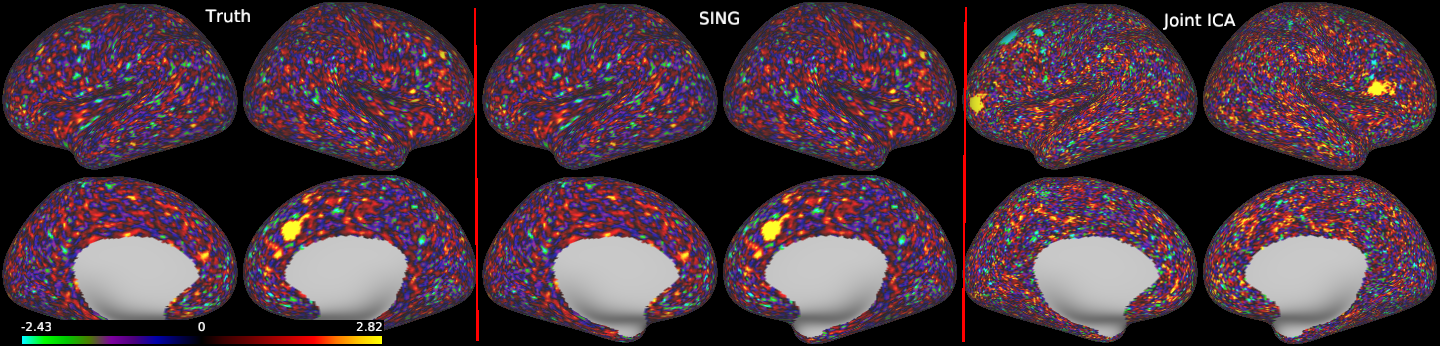}
     \includegraphics[width = \textwidth]{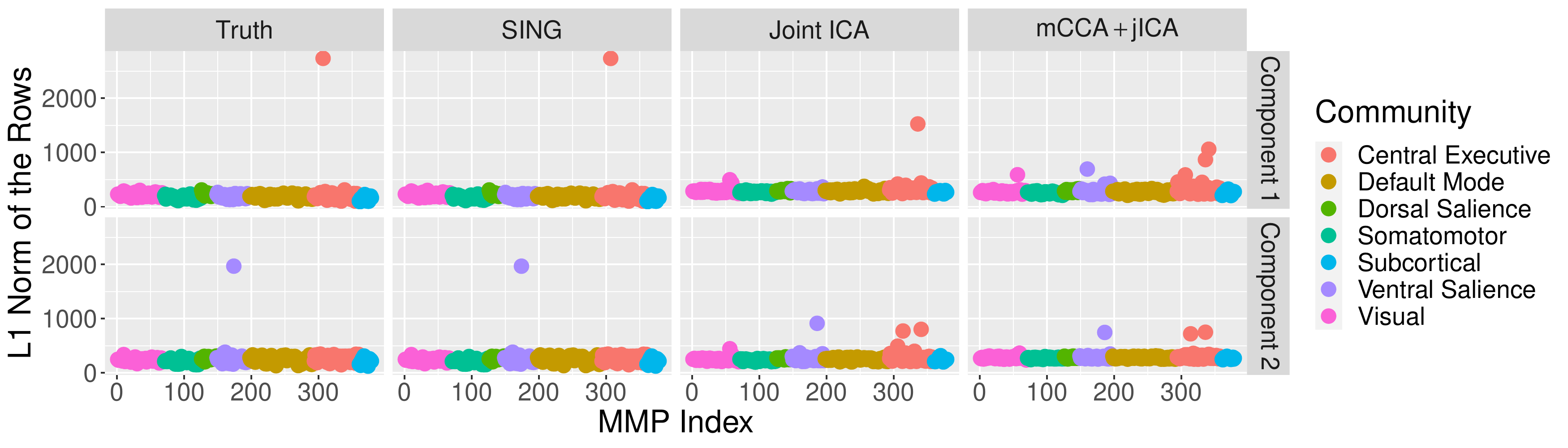}
    \caption{Simulation setting 2 of Section~\ref{sec:sim_large}. Comparison of loadings of the joint components. \underline{Top and middle rows:} Joint loadings $\bS_{Jx}$ and $\hatbS_{Jx}$. A row of the joint loadings matrix is visualized on the cortical surface using four views corresponding to the lateral left hemisphere, lateral right, medial right, and medial left views. The gray area denotes the medial wall and does not contain gray matter. SING recovers the true loadings. Joint ICA and jICA+mCCA are inaccurate because these methods use PCA for dimension reduction, and the joint components simulated here have low variance. Both the second Joint ICA component loadings and second mCCA+jICA component loadings spuriously associate features from the individual non-Gaussian components (see Figure~\ref{fig:simspur2} for comparison with individual components). \underline{Bottom row}: L1 norms of the rows of the network matrice s formed from $\bS_{Jy}$ and $\hatbS_{Jy}$. The $p \times p$ representations are in Figure~\ref{fig:Ycomponent1}. The plots of Truth and SING look very similar, revealing accurate estimation of joint loadings. Joint ICA and mCCA+JICA loadings do not correspond to the truth. Additional details in Table~\ref{tab:sim_large}.}
    \label{fig:ComponentsCifti}
\end{figure}

Figure~\ref{fig:ComponentsCifti} compares true $\bS_{Jx}$ and $\bS_{Jy}$ with $\hatbS_{Jx}$ and $\hatbS_{Jy}$ estimated by SING with large $\rho$, Joint ICA, and mCCA+jICA. Panel a in Figure~\ref{fig:ComponentsCifti} depicts the first row of $\bS_{Jx}$ plotted on the cortical surface. $\hatbS_{Jx}$ from SING is visually indistinguishable from the true components (panels b versus a, f versus e). In contrast, neither Joint ICA (panels c and g) nor mCCA+jICA (panels d and h) capture the true loadings. In Joint ICA component 2, the features that appear in Figure~\ref{fig:ComponentsCifti} are found in the true individual loadings $\bS_{Ix}$, but were spuriously estimated as part of the joint structure, as depicted in Figure~\ref{fig:simspur2}.  To create the scatterplots in the bottom row in Figure~\ref{fig:ComponentsCifti}, we first constructed the $379 \times 379$ loadings matrices from the vectorized loadings $\hatbS_{Jy}$, as depicted in Figure~\ref{fig:Ycomponent1}. We then summed the absolute values of the rows of the loadings matrices to simplify the presentation in Figure \ref{fig:ComponentsCifti}. One node is prominent for each of the true component loadings, and SING accurately captures the same nodes. However, prominent nodes in Joint ICA and mCCA+jICA joint loadings do not coincide with the truth for either component.

In Simulation Setting 2, the true components $\bS_x$ and $\bS_y$ do not contain exact zeros. Sparse brain activation maps and networks may be more interpretable. We thus also considered a simulation where the true $\bS_x$ and $\bS_y$ are exactly sparse (Simulation Setting 3 in Appendix~\ref{sec:sim_sparse}), and obtained the same qualitative conclusions.

\section{Analysis of task and resting-state data from the Human Connectome Project}\label{sec:data}

\subsection{Data and Methods}\label{sec:hcpdata}

We applied SING to estimate the shared structure in the working memory task (task maps) and resting state correlations (rs correlations) from the Human Connectome Project. In the SING decomposition, a loading for a joint component in the working memory dataset represents the importance of a vertex (spatial location) to the component, and can be interpreted in a manner similar to a task activation map. A loading in the rs correlations represents the importance of an edge (pair of regions) to the component, and the loadings form a symmetric matrix that can be interpreted in a manner similar to a connectivity matrix. A subject score captures the importance of the linked loadings to the individual, such that a large score indicates the associated component is prominent in that subject. If the subject scores of a joint component are related to fluid intelligence, then the loadings correspond to the locations and edges associated with fluid intelligence. We summarize the data here with details provided in the Appendix~\ref{appendix:HCP}. We used the statistical parametric maps created by the HCP from the working memory task. An example subject is depicted in Figure~\ref{fig:example}. The data are the z-statistics at $p_x =59,\!412$ vertices (cortical locations) from the contrast of the 2-back and 0-back memory tasks and represent an estimate of the cortical regions that are engaged in working memory. Working memory involves the manipulation of temporarily stored information. In the 2-back working memory task, a participant presses a clicker only when she/he sees a picture that was viewed prior to the previous picture, whereas as in a 0-back task, a participant presses a clicker if they view a pre-specified picture at any point during the sequence. Data are vectorized for input to SING. For rs correlations, we used the ICA-FIX preprocessed rs-fMRI data from subjects with four rs-fMRI scans \citep{glasser2013minimal}. Time courses for each vertex were standardized (zero mean and variance one) prior to averaging the vertices within a region. For each subject, correlation matrices for each run were created, Fisher-transformed, and then averaged. Example subjects are shown in Figure~\ref{fig:example2}. The lower triangular values were extracted to form a vector of length $p_y =71,\!361$ for each subject.

There were $n=996$ subjects with task maps and rs correlations. We standardized each feature (across subjects), centered each subject (across features), and iterated this until centering each subject had a negligible impact on the variance of the features (here, 6 iterations, such that the variance across rows for each feature equaled one, the mean across rows equaled zero, and the mean across columns equaled zero). Hence the rank of each dataset is 995.

We initially estimated 995 components in each dataset. However, there is a well-known problem of local minima in fitting ICA models due to non-convexity that is even more pronounced for higher dimensions \citep{risk2014evaluation}. To reduce sensitivity to initializations, we utilized 100 initializations and identified the estimate with the largest objective function value, which is our estimate of the argmax. We matched each initialization by absolute correlation with the argmax. Repeating this across initializations, this created a 995$\times$100 matrix. We then retained all non-Gaussian components in the argmax in which the absolute correlation was $>0.95$ in 75\% of the initializations. This resulted in 156 non-Gaussian components in the task maps and 611 components in the rs correlations. We then applied the joint rank test with $\alpha=0.01$, which resulted in 30 components (min correlation between matched columns = 0.17, max correlation = 0.34). We chose $\rho$ to result in correlations between columns of $\hatbM_{Jx}$ and $\hatbM_{Jy}$ $\ge 0.99$. Here, $\rho$ was set to the sum of the JB values of the joint components divided by ten.
We then repeated this estimation scheme for a second set of 100 initializations and compared the resulting components between the two batches. In the second batch, the first stage resulted in 155 task map components and 609 rs correlation components. The joint rank test resulted in 32 components. To examine the impact of $\alpha$, we also conducted the permutation test with $\alpha=0.05$ and obtained 35 and 40 components, respectively. Plots of the matched correlations are available in Figure~\ref{fig:hcp_jointrank}.  
We applied SING to the two batches with components from $\alpha=0.01$. We matched the columns of the joint mixing matrices from the two batches and retained all joint components with absolute correlations $>0.95$, resulting in 26 joint components. We performed the same matching procedure to obtain individual components from each dataset, resulting in 103 and 553 individual components.

Each initialization of the separate LNGCA took approximately 2.2 hours (range 1.6-2.9) in the task maps and 1.5 hours (range 1.0-2.0) for the rs correlations on a cluster with 2.4 GHz Xeon CPUs. For each batch, the greedy matching algorithm took $<$1 second; the permutation test took $<$1 minute; the SING algorithm took 17 hours to converge with tolerance 1e-06.

To explore the biological meaning in joint subject scores, we examined the relationship between fluid intelligence and joint subject scores. Subjects' abstraction and mental flexibility were measured in the HCP behavioral data using the Penn Matrix Reasoning Task A (PMAT24\_A\_CR), hereafter fluid intelligence. We created a multiple linear regression model predicting fluid intelligence from the 26 joint subject scores. 

We also estimated joint components using Joint ICA and mCCA+jICA. We again used the JB statistic as our measure of non-Gaussianity for both methods. In Joint ICA, the number of components was set to 26, and 100 initializations were used to find the argmax. In mCCA+jICA, we used 135 components in the initial PCA for the working memory, and 135 components in the initial PCA for the rs correlations. This is the number of components used in Cohort 1 in the mCCA+jICA in \cite{lerman2017multimodal}. We then estimated 26 canonical variables for input to the joint ICA step with 100 initializations.

\subsection{Results}\label{sec:HCPResults}

\begin{figure}
    \centering
    \includegraphics[width=0.64\textwidth]{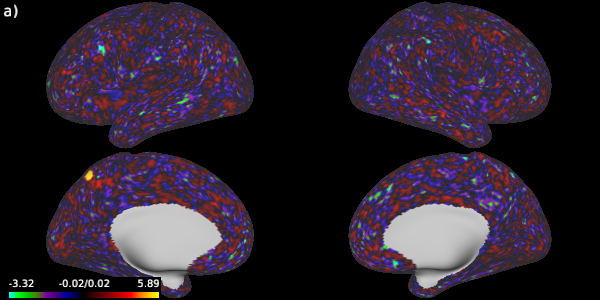}
      \includegraphics[width=0.32\textwidth]{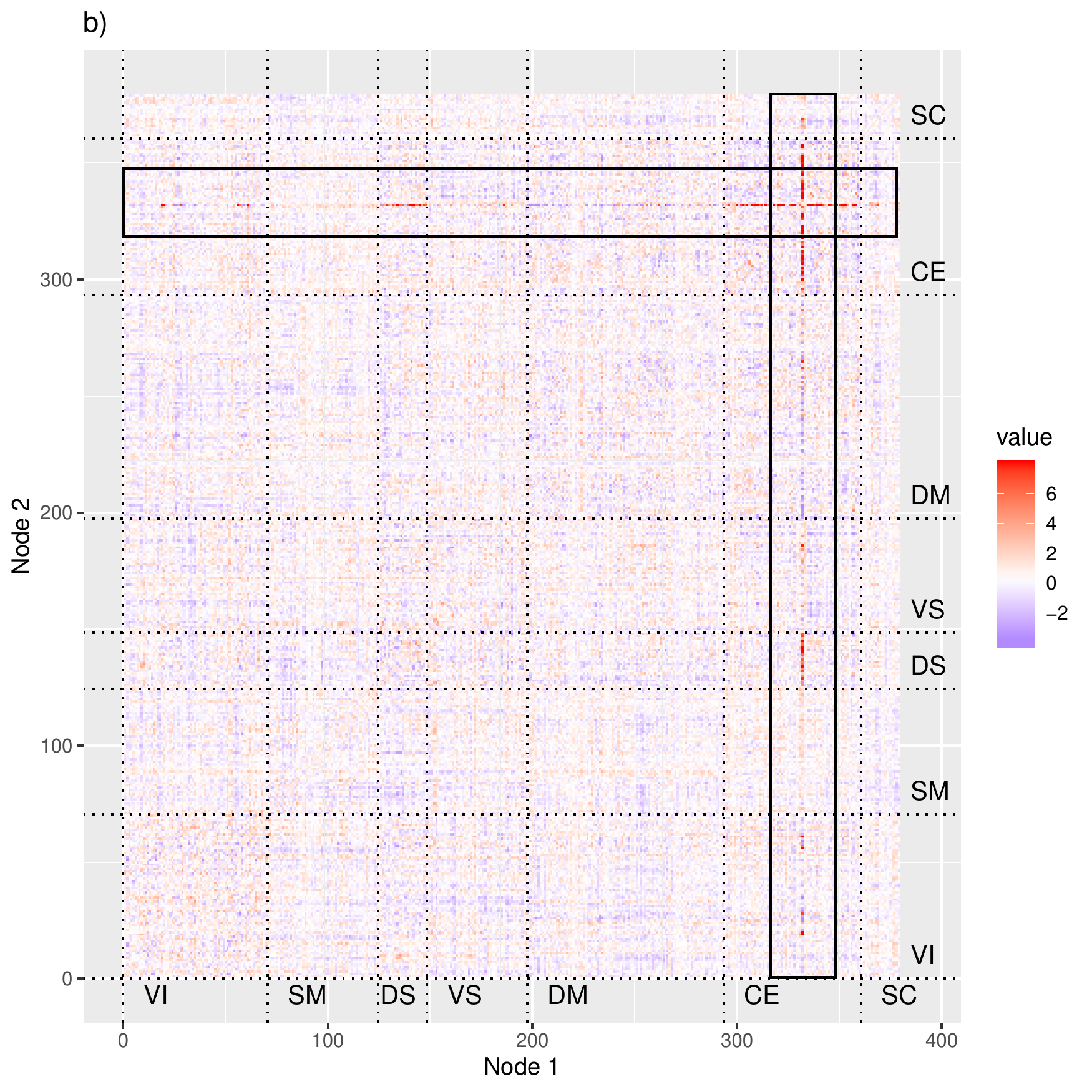}
      
    \includegraphics[width=0.32\textwidth]{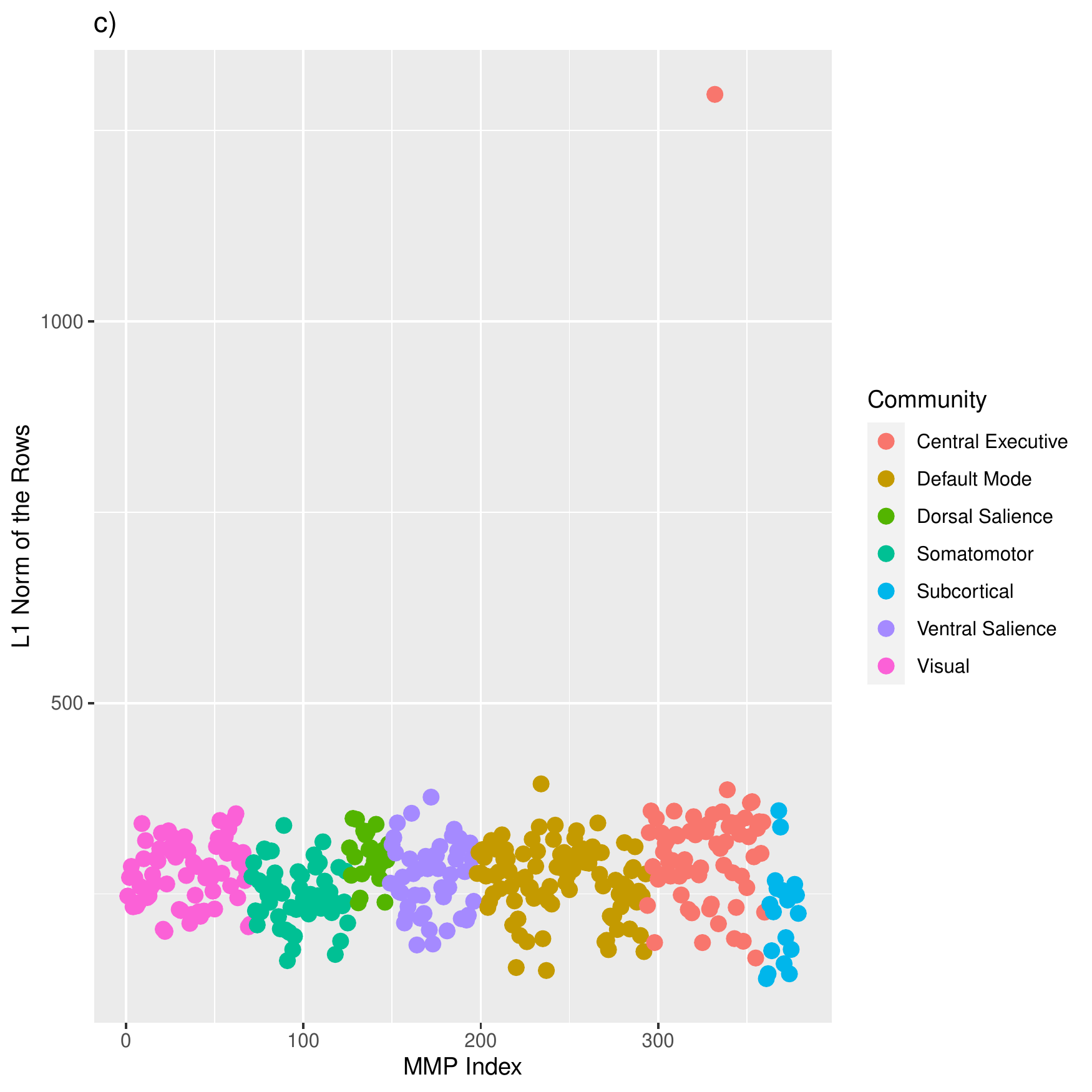}
    \includegraphics[width=0.64\textwidth]{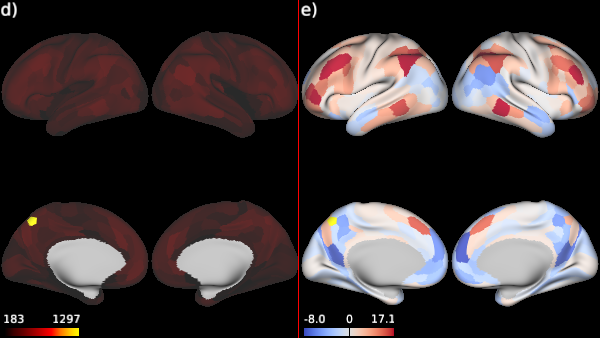}
    \caption{SING component corresponding to subject scores most strongly related to fluid intelligence. a) Joint component loadings from the working memory task. b) Joint component loadings from the rs correlations. VI: visual; SM: somatomotor; DS: dorsal salience; VS: ventral salience; DM: default mode; CE: central executive; SC: subcortical and cerebellum. c) L1 norms of the edges in each region (equal to the sum of the absolute values of the rows of b). d) L1 norms for the regions visualized on the cortical surface. e) Loadings for the row of b) corresponding to the prominent region (L\_7Pm\_ROI) plotted on the cortical surface.}\label{fig:hcpsing}
\end{figure}

Our results indicate there is joint information in the working memory task maps and rs correlations. The rows of $\hatbS_{Jx}$ and $\hatbS_{Jy}$ exhibit strong spatial correspondence.  We provide an example of this by a detailed examination of the joint scores that were most strongly related to fluid intelligence in the multiple regression. In the results that follow, the signs of the components were chosen to result in positive skewness. Component 24 ($t=5.13$, $p<1e-06$ uncorrected) has a small patch of cortex with large task map loadings (Figure \ref{fig:hcpsing}a). The loadings of the rs correlations are prominent in edges corresponding to a single node, visible as a red cross in Figure \ref{fig:hcpsing}b. We calculate the L1 norm of each row, reducing the 379$\times$379 matrix of loadings to 379 points, which are colored by their community membership from \cite{akiki2018determining} in Figure \ref{fig:hcpsing}c. This shows a single prominent node in the central executive network. We then plot the L1 norms for all nodes on their corresponding locations on the cortical surface (Figure \ref{fig:hcpsing}d). The prominent node in this rs correlation component strongly coincides with the spatial locations of the most activated vertices in the task map component (Figure \ref{fig:hcpsing}a and d). This node is called the left medial area 7P region of interest (L\_7Pm\_ROI) and is a member of the central executive community, which is a network involved in cognition and memory. In \cite{glasser2016multi}, L\_7Pm\_ROI is described as more activated in the working memory task relative to its neighboring parcels. Also, Figure \ref{fig:hcpsing}b) shows it has strong connections with other nodes in the central executive as well as nodes in the dorsal salience network, and some opposite connections with the default mode network. In Figure \ref{fig:hcpsing}e), we plot the row of Figure \ref{fig:hcpsing}b) corresponding to L\_7Pm\_ROI, where the red areas indicate positive loadings, which occur in nodes of the central executive and dorsal salience network, while blue areas indicate negative loadings, which occur in the default mode network (see \citealt{akiki2018determining}). The salience network is thought to be involved in switching between the default mode network and central executive \citep{goulden2014salience}. The default mode network tends be prominent when subjects are not focusing on the external world, including day dreaming and self reflection, and the opposite loadings relative to central executive suggest less activity in this region for the subjects with higher activity in L\_7Pm\_ROI.

\begin{figure}
    \centering
    \includegraphics[width=\textwidth]{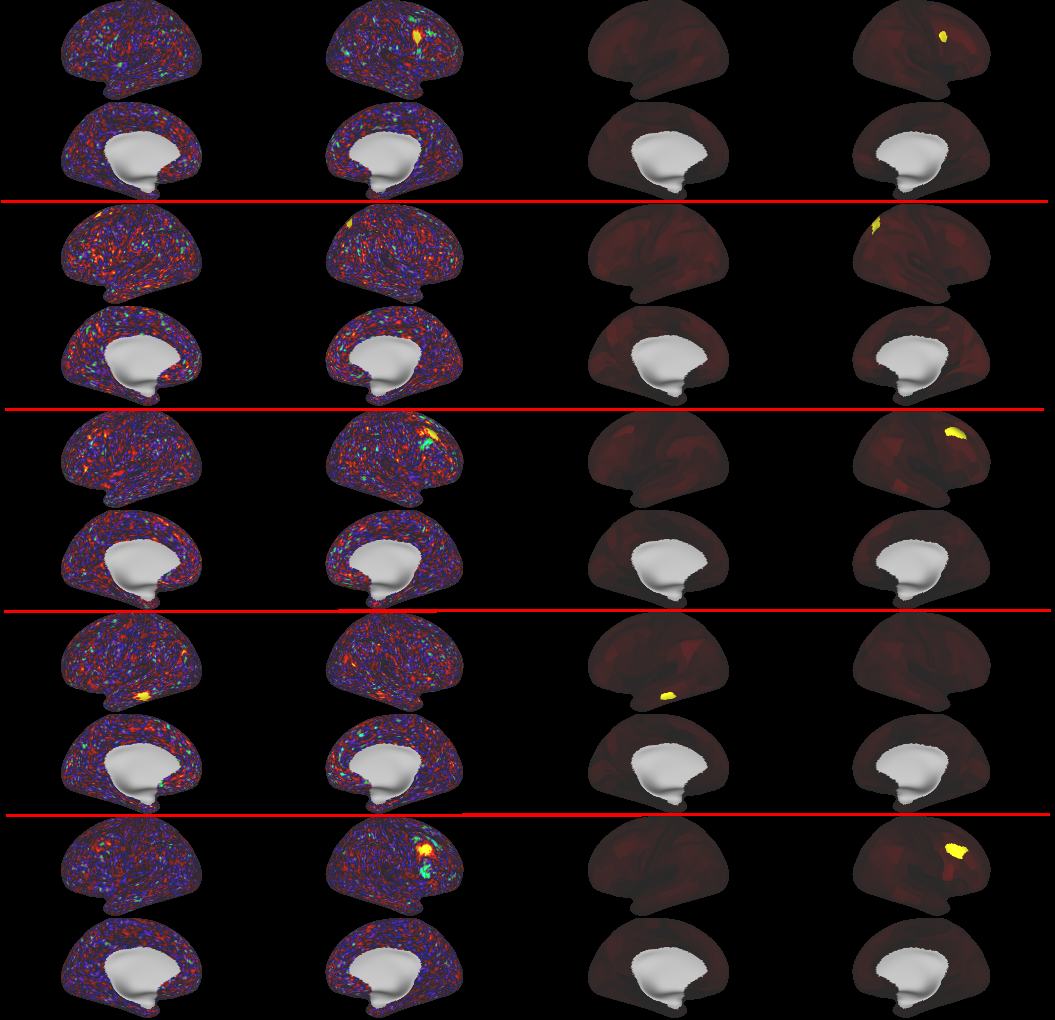}
    \caption{Five components from SING. Each component has four views (clockwise: lateral left, lateral right, medial right, medial left) from the task map component (left two columns) and four views from the L1 norms of the edges for each node in the symmetric loadings matrix from the rs correlations (right two columns). These five components were selected because the areas of activation are visible on the inflated surface in a common orientation. In all images, the colorbar is set with absolute percentage [2, 99.9].}
    \label{fig:SING_5comps}
\end{figure}

The strong spatial correspondence applies to the other joint components, five of which are displayed in Figure \ref{fig:SING_5comps}. A visual examination of the 26 non-Gaussian components from the task maps and the L1 norms in the rs correlations suggests some spatial correspondence in all but one component. The joint structure represents a small amount of the total variation in each dataset: $R_{Jx}^2=0.023$ and $R_{Jy}^2=0.024$ for the task maps and rs correlations, respectively, which underscores how our method finds information that is distinct from variance-based approaches. This spatial correspondence is learned in an unsupervised manner, as no information about the spatial locations is involved in the model estimation. 

To gain insight into the regions extracted by SING, we created composite images of the joint loadings for the task and rs correlations. The sum of the absolute values of the loadings across components shares some similarities to a subset of the regions highlighted in the task activation maps, in particular highlighting lateral regions of the prefrontal cortex, medial regions near left and right 7Pm\_ROI, and some portions of the temporal lobe (Figure~\ref{fig:composite}). This suggests that in a different task, the regions highlighted in the rs correlation loadings would be related to the specific task, rather than an underlying similarity between fMRI datasets. To examine whether the subject scores differed greatly in magnitude between datasets, we calculated $\widehat{\bD}_x$ and $\widehat{\bD}_y$ (corresponding to working memory and rs correlations, respectively). A joint component with subject scores that are very small in scale relative to other components may represent a component that holds negligible joint structure. The diagonal elements of $\widehat{\bD}_x$ ranged from 0.86 to 1.03, and those in $\widehat{\bD}_y$ ranged from 0.71 to 1.28. This provides evidence that the joint components represent non-trivial structure in all cases. Additionally, the ratio of the diagonal elements of $\widehat{\bD}_x$ to $\widehat{\bD}_y$ varied from 0.71 to 1.45. Taken together, these results indicate SING can be a powerful tool to discover meaningful shared structure between datasets. 

To examine the relative importance of the joint versus individual information within each dataset, we compared the L2 norms of the columns of $\hatbM_{Ix}$ to the diagonal elements of $\widehat{\bD}_x$, and similarly for $\hatbM_{Iy}$ and $\widehat{\bD}_y$. For the working memory task, the norms of the 103 columns of $\hatbM_{Ix}$ ranged from 0.97 to 1.99 with median 1.12, compared to the 26 diagonal elements of $\widehat{\bD}_x$ ranging from 0.86 to 1.03 with median 0.94. For the rs correlations, the norms of the 553 columns of $\hatbM_{Iy}$ ranged from 0.47 to 2.20 with median 1.06, compared to the 26 diagonal elements from $\widehat{\bD}_y$ ranging from 0.71 to 1.28 with median 0.91. Thus in both datasets, the contribution of a typical individual component tended to be larger than a typical joint component, although there was wide overlap in the ranges.

The estimated components in Joint ICA and mCCA+jICA differ greatly from SING, and the shared structure of their joint components is harder to adjudicate. In Joint ICA, one modality tends to contain clear structure and the other modality contains less structure, as shown in Figure~\ref{fig:jointICA5}. In particular, there is little spatial correspondence between components. In mCCA+jICA, the correlation between the subject scores from the same component ranged from 0.70 (component 1) to 0.31 (component 13), with a trend in decreasing correlations with higher component number (correlation between component correlations and component number = -0.68). In mCCA+jICA, there tends to be spatial correspondence in early components, but later components exhibit a lack of spatial correspondence. See Figure~\ref{fig:mCCA5} for components 1, 7, 13, 19, and 25.  

\begin{figure}
     \centering
     \includegraphics[width=\textwidth]{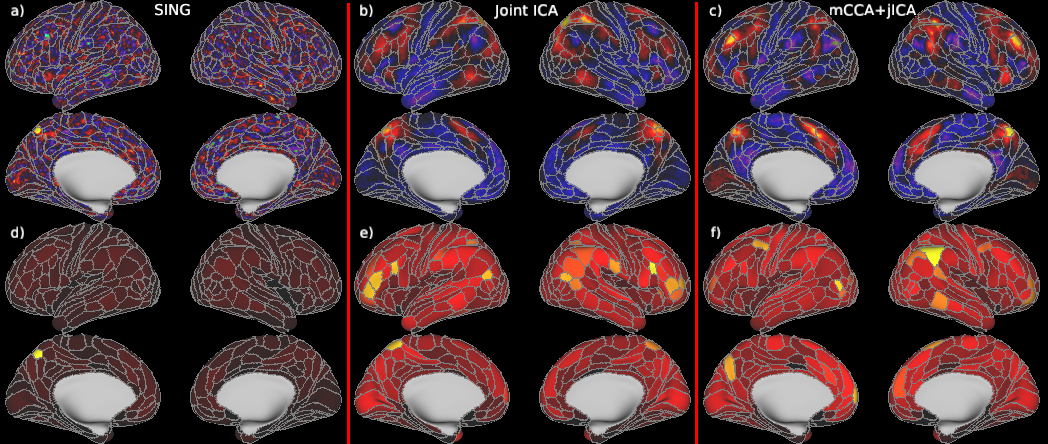}
\caption{Loadings from a component strongly related to fluid intelligence in SING, Joint ICA, and mCCA+jICA. Top: working memory loadings. Bottom: l1 norms of the rows of the rs correlation loadings. In all methods, parcel L\_7Pm\_ROI has large loadings in the working memory task (area in yellow in panel a, also visible in b and c). This area is also prominent in the rs correlations in SING (panel d), but not in Joint ICA (panel e) nor mCCA+jICA (panel f). Panels a) and d) correspond to the SING component depicted in Figure \ref{fig:hcpsing}. In all images, the colorbar is set with absolute percentage [2, 99.9].}
     \label{fig:CompareComponents_fluidintelligence_nocolorbars}
 \end{figure}

We compared a component from each method that was strongly related to fluid intelligence. Figures  \ref{fig:CompareComponents_fluidintelligence_nocolorbars}a and d show the SING component from Figure \ref{fig:hcpsing}. The area highlighted in the working memory loadings in SING (Figure \ref{fig:CompareComponents_fluidintelligence_nocolorbars}a, yellow in the medial view of the left hemisphere) is also prominent in the working memory loadings in Joint ICA and mCCA+jICA (Figure \ref{fig:CompareComponents_fluidintelligence_nocolorbars}b and c, respectively). However, the corresponding region of interest in the rs correlation loadings is not visible in Joint ICA or mCCA+jICA (Figure \ref{fig:CompareComponents_fluidintelligence_nocolorbars}e and f, respectively). Also notable is that the contralateral parcel R\_7Pm\_ROI in SING is highlighted in component 11 (not shown), again with high spatial correspondence between working memory and rs correlations, and is also associated with fluid intelligence ($t=4.5$, $p<1e-05$). In Joint ICA, the subject scores corresponding to the loadings depicted in Figure \ref{fig:CompareComponents_fluidintelligence_nocolorbars}b and e are strongly related to fluid intelligence ($t=5.10$, $p<1e-06$), but there is little evidence of spatial correspondence. In mCCA+jICA, the subjects scores corresponding to panels c and f are most strongly related to fluid intelligence ($t=11.34$, $p<1e-16$, and $t=7.62$, $p<1e-13$, respectively; correlation between scores: $\rho=0.46$). In panel f, the yellow area in the lateral view of the right hemisphere of the L1 norms of the rs correlation loadings corresponds to R\_PFm\_ROI,  which is a major node in the task positive network. This region is visible in red and orange in the working memory loadings (panel c). However, the prominent yellow areas in the working memory loadings roughly correspond to R\_7Pm\_ROIs and L\_7Pm\_ROI, which is the region discussed in the SING component, and R\_p9-46v\_ROI and L\_p9-46v\_ROI, which is in the dorsolateral prefrontal cortex. These areas are not apparent in the rs correlation loadings in panel f. The observed spatial correspondence in SING, but not in Joint ICA or mCCA+jICA, is consistent with our large-scale simulation (Figure \ref{fig:ComponentsCifti}). Additional discussion of joint loadings is in Appendix~\ref{appendix:HCP} and Figure~\ref{fig:hcpjointica}. 
 
\section{Discussion}

We propose Simultaneous Non-Gaussian component analysis (SING) for extracting shared structure from two datasets, which reveals a different view of human brain function compared to two existing methods. In simulations, our algorithm extracts shared structure, whereas Joint ICA and mCCA+jICA discard this structure and/or incorrectly associate structure unique to a dataset. In our application to the Human Connectome Project, each component identifies spatial locations of high activation in a working memory task that are linked to a spatially coinciding region in the resting-state correlations. For the regions that contribute to working memory, SING reveals those locations that also have a large contribution to brain activity in the resting state. The spatial correspondence is learned from the model, and in itself is an important biological discovery.

Our results may shed light into the regional-specialization paradigm and network paradigm of the brain: a specialized region in the task fMRI is associated with the same region in the rs correlations plus a picture of its connections to other regions. It has been hypothesized that the salience network mediates between central executive (which involves working memory) and default-mode networks \citep{bressler2010large}. We provided an illustration of the meaning in the subject scores and loadings by examining the component most strongly related to fluid intelligence. The locations highlighted in the working memory task coincided with the L\_7Pm\_ROI in the resting-state correlation loadings, which is in the central executive network.  The rs correlation loadings in SING also revealed that this region has positive connections with regions in the dorsal salience network and negative connections with the default mode network, lending support to the multi-network switching hypothesis. The region 7Pm\_ROI was also prominent in the working memory loadings in Joint ICA and mCCA+jICA; however, the resting-state correlation loadings did not have spatial correspondence.  In a related study, \cite{lerman2017multimodal} examined four modalities from the HCP dataset: the 2-Back working memory activation maps, activation maps from a relationship processing task, cortical thickness, and rs correlations. They examined how subject scores were related to a composite measure of cognition, which was based on four variables including the measure of fluid intelligence examined here. They identified one component significantly related to cognition for all four modalities. The loadings highlighted regions of the visual system in both tasks, as well as a small region near 7Pm\_ROI in the 2-Back task. The resting-state correlations indicated modular community organization with the visual system, dorsal attention, cingulo-opercular, while cortical thickness loadings were prominent in the insula. The authors note that a limitation of their approach is that ICA-based methods may not guarantee a perfect decomposition and separation of sources. We agree this is a limitation of current approaches, and we suggest SING is an improvement. A limitation of our approach is that we only considered two datasets, and an important avenue for future research is the extension to multiple datasets that may contain partially shared structure. For each subject, our rs correlations were generated by averaging correlations from four scans. An interesting extension would be to develop a hierarchical model that can leverage longitudinal measurements on the same modality.

Our approach results in components with smaller brain regions of high activation compared to Joint ICA and mCCA+jICA. Previous studies applying ICA to subject-by-voxel data have used an empirical approach to selecting the number of components. In \cite{willette2014prognostic}, 20 components resulted in ``poor specificity'' while a larger number of components produced ``fragmentary'' ICs. We speculate that the smaller areas of cortex in SING may result in improved specificity in the sense that the subject scores are localized to smaller brain regions; in contrast, approaches with larger brain regions may aggregate subregions that perform different functions, decreasing the specificity of the associated subject scores.

There are some shortcomings of the current work that should be investigated in future research. Local optima are a challenge in ICA and LNGCA because the Stiefel manifold is a non-convex space, and the choice of non-Gaussian measure may also be non-convex, as in the JB statistic. In practice, we have found that it is more challenging to find the argmax when estimating a large number of components, as in our data example, than when estimating components from an initial PCA step, although issues with local minima do also occur in the latter case \citep{risk2014evaluation}. Additionally, the number of components may be sensitive to the choice of $\alpha$ in data applications due to the lack of a clear gap between the correlations of joint and individual components. Hence, we recommend using two sets of initializations and retaining reliable components, as described in Section \ref{sec:hcpdata}, but acknowledge this approach may discard some non-Gaussian information.   

We examined the relationship between a working memory task and rs-fMRI. Figure~\ref{fig:composite} suggests a substantial portion of the locations with large loadings in the task are associated with working memory activation, and similarly for the loadings in the rs correlations.  An important avenue for future research is to examine which regions of the brain are prominent when integrating rs-fMRI with other tasks. For example, areas of the motor cortex may be highlighted in rs-fMRI when analyzing shared structure between a motor task and resting-state data. While we used Pearson correlations as our estimate of functional connectivity, \cite{mohanty2020rethinking} suggest a composite measure of functional connectivity including Pearson correlation, coherence, mutual information, and dissimilarity measures. SING could be investigated as a method to integrate multiple measures. Another important avenue for future research is a sparse method that produces exact zeros in the loadings, for example, by using the L1-norm to measure non-Gaussianity of loadings rather than the JB statistic employed here. This may improve interpretability.

\section*{Acknowledgements}
IG was supported in part by NSF DMS-1712943. Data were provided [in part] by the Human Connectome Project, WU-Minn Consortium (Principal Investigators: David Van Essen and Kamil Ugurbil; 1U54MH091657) funded by the 16 NIH Institutes and Centers that support the NIH Blueprint for Neuroscience Research; and by the McDonnell Center for Systems Neuroscience at Washington University.

\section*{R Code and Simulated Dataset}
R code to replicate the simulations is available at \url{https://github.com/irinagain/SING}.

%%%%%%%%%%%%%%%%%%%%%%%%%%%%%%%%%%%%%%%%%%%%%%
%% Example with multiple Appendixes:        %%
%%%%%%%%%%%%%%%%%%%%%%%%%%%%%%%%%%%%%%%%%%%%%%
\begin{appendix}

\section{Initializing the curvilinear algorithm}\label{appendix:InitialAlgorithm}

We generate initial estimates for SING by fitting LNGCA with the JB statistic separately to each dataset, where non-Gaussianity is maximized across features. 
We match the columns of $\widehat \bM_x$ and $\widehat \bM_y$ using the greedy algorithm with $r_J$ estimated using the permutation test (Section \ref{sec:rJ}). Denote these permuted estimates with first $r_J$ matched columns as $\widehat \bM_{x}^{(0)}$ and $\widehat \bM_{y}^{(0)}$. Then the input to SING is $\widehat \bU_x^{(0)} = \hatbL_x \hatbM_{x}^{(0)}$, where $\hatbL_x$ is the whitening matrix for $\bX$, and similarly define $\widehat \bU_y^{(0)}$. 

\section{Derivation of Algorithm 1.}\label{appendix:algorithm1}

In this section we derive the modification of the curvilinear search algorithm \citep{Wen:2012ga} for problem~\eqref{eq:joint_app3}. We start by deriving the gradient of the objective function with respect to $\bu_{xl}$ and $\bu_{yl}$, respectively. For $\alpha \in [0,1]$,
\begin{align*}
f(\bu_{xl}^{\top}\bX_w) &= \alpha\Big\{\frac1{p_x}\sum_{j=1}^{p_x}(\bu_{xl}^{\top}\bX_{wj})^3\Big\}^2 + (1-\alpha)\Big\{\frac1{p_x}\sum_{j=1}^{p_x}(\bu_{xl}^{\top}\bX_{wj})^4-3\Big\}^2\\
&= \alpha \gamma^2(\bu_{xl}^{\top}\bX_{wj}) + (1-\alpha)\kappa^2(\bu_{xl}^{\top}\bX_{wj}).
\end{align*}

Taking the gradient with respect to $\bu_{xl}$ gives
\begin{align*}
    t_{\alpha}(\bu_{xl}) = \frac{\partial f(\bu_{xl}^{\top}\bX_w)}{\partial \bu_{xl}} &= 6\alpha \gamma(\bu_{xl}^{\top}\bX_{wj})\sum_{j=1}^p \{\bX_{wj}(\bu_{xl}^{\top}\bX_{wj})^2\} \\
    &\quad + 8 (1-\alpha)\kappa(\bu_{xl}^{\top}\bX_{wj})\sum_{j=1}^p \{\bX_{wj}(\bu_{xl}^{\top}\bX_{wj})^3\}.
\end{align*}
The form for $\bu_{yl}$ is analogous.

Next, for $l\leq r_J$, consider the distance
\begin{align*}
d(\widehat \bL_x^{-1}\bu_{xl}, \hL_y^{-1}\bu_{yl}) = \left\|\frac{\hL_x^{-1}\bu_{xl}\bu_{xl}^{\top}\hL_x^{-1}}{\bu_{xl}^{\top}\hL_x^{-2}\bu_{xl}}-\frac{\hL_y^{-1}\bu_{yl}\bu_{yl}^{\top}\hL_y^{-1}}{\bu_{yl}^{\top}\hL_y^{-2}\bu_{yl}}\right\|_F^2 = 2 - 2\left(\frac{\bu_{xl}^{\top}\hL_x^{-1}\hL_y^{-1}\bu_{yl}}{\sqrt{\bu_{xl}^{\top}\hL_x^{-2}\bu_{xl}\bu_{yl}^{\top}\hL_y^{-2}\bu_{yl}}}\right)^2.
\end{align*}
Let $\ba_{yl} = (\widehat \bL_y^{-1}\bu_{yl})/\sqrt{\bu_{yl}^{\top}\widehat \bL_y^{-2}\bu_{yl}}$, then
$$
d(\widehat \bL_x^{-1}\bu_{xl}, \hL_y^{-1}\bu_{yl}) = d(\widehat \bL_x^{-1}\bu_{xl}, \ba_{yl}) = 2 - 2\left(\frac{\bu_{xl}^{\top}\hL_x^{-1}\ba_{yl}}{\sqrt{\bu_{xl}^{\top}\hL_x^{-2}\bu_{xl}}}\right)^2,
$$
and calculating the gradient with respect to $\bu_{xl}$ gives
\begin{equation*}
\frac{\partial d(\widehat \bL_x^{-1}\bu_{xl}, \ba_{yl})}{\partial \bu_{xl}}  =   -4\left(\frac{\bu_{xl}^{\top}\hL_x^{-1}\ba_{yl}}{\sqrt{\bu_{xl}^{\top}\hL_x^{-2}\bu_{xl}}}\right)\left(\frac{\hL_x^{-1}\ba_{yl}}{\sqrt{\bu_{xl}^{\top}\hL_x^{-2}\bu_{xl}}} - \frac{\left(\bu_{xl}^{\top}\hL_x^{-1}\ba_{yl}\right)\hL_x^{-2}\bu_{xl}}{(\bu_{xl}^{\top}\hL_x^{-2}\bu_{xl})^{3/2}}\right).
\end{equation*}

Combining the above two displays, the gradient of objective function in~\eqref{eq:joint_app3} with respect to $\bu_{xl}$ for $l\leq r_J$ is equal to
\begin{align*}
g_x(\bu_{xl}, \bu_{yl}) = - t_{\alpha}(\bu_{xl}) - 4\rho\left(\frac{\bu_{xl}^{\top}L_x^{-1}\ba_{yl}}{\sqrt{\bu_{xl}^{\top}\hL_x^{-2}\bu_{xl}}}\right)\left(\frac{\hL_x^{-1}\ba_{yl}}{\sqrt{\bu_{xl}^{\top}\hL_x^{-2}\bu_{xl}}} - \frac{\left(\bu_{xl}^{\top}\hL_x^{-1}\ba_{yl}\right)\hL_x^{-2}\bu_{xl}}{(\bu_{xl}^{\top}\hL_x^{-2}\bu_{xl})^{3/2}}\right),
\end{align*}
and for $l > r_J$ is equal to
$
g_x(\bu_{xl}) = - t_{\alpha}(\bu_{xl}).
$

Let $G_x(\bU_x, \bU_y) = [g_x(\bu_{x1}, 
\bu_{y1})\dots g_x(\bu_{xr_J}, \bu_{yr_J})\ g_x(\bu_{x(r_{J}+1)})\dots g_x(\bu_{xr_x})]\in \R^{n\times r_x}$. Then given $\tau >0$, the curvilinear algorithm proceeds by generating updates
\begin{align*}
\bV_x(\tau) = \bU_x\Big(\bI-\frac{\tau}2\bW_x\Big)\Big(\bI+\frac{\tau}2\bW_x\Big)^{-1}
\end{align*}
based on the skew-symmetric matrix $\bW_x = \bU_x^{\top}G_x(\bU_x, \bU_y)^{\top} - G_x(\bU_x, \bU_y)\bU_x$. The updates for $\bU_y$ are similar:
\begin{align*}
\bV_y(\tau) = \bU_y\Big(\bI-\frac{\tau}2\bW_y\Big)\Big(\bI+\frac{\tau}2\bW_y\Big)^{-1}
\end{align*}
based on the skew-symmetric matrix $\bW_y = \bU_y^{\top}G_y(\bU_x,\bU_y)^{\top} - G_y(\bU_x,\bU_y)\bU_y$. 

The original curvilinear search algorithm is described with respect to only one orthogonal constraint. One can alternate updates with respect to $\bU_x$ and $\bU_y$ (thus considering only one constraint at a time), or jointly update $\bU_x$ and $\bU_y$. We found that the joint update step leads to considerably faster convergence, which is the approach we take. In practice, we update $\tau$ in each step $k$, which we denote as $\tau_k$. We set $\tau_k = 0.01(0.8)^h$, where $h$ is the minimal non-negative integer such that the objective function value evaluated at $\bV_x(\tau_k), \bV_y(\tau_k)$ is smaller than the objective function value at the current $\bU_x$, $\bU_y$.

\section{Signal rank estimation alternatives}\label{appendix:rank}
In this section we describe a new sequential permutation test that can be used for estimation of signal rank in LNGCA.

We conduct a sequence of tests for $r=1,\dots,n-1$. Suppose $r$ components are estimated. The following tests if the $r$th component is non-Gaussian.
\begin{enumerate}
    \item Let $f(\hatbS_1),\dots,f(\hatbS_r)$ denote the JB-statistics for each of the $r$ estimated components, where the statistics are in descending order.
    \item Let $\hatbS_{1:(r-1)}$ denote the first $r-1$ ordered non-Gaussian components. Define $\bX_{resid,r-1} = \bX_c - \bX_c \widehat{\bS}_{1:(r-1)}^\top \widehat{\bS}_{1:(r-1)}$.
    \item  For $t=1,\dots,T$: 
    \begin{enumerate} 
    \item Randomly select $n-r+1$ subjects.
    \item Permute the columns of each row retained from $\bX_{resid,r-1}$. Then the features for each subject are mismatched, which simulates the null distribution of no non-Gaussian features shared across subjects.
    \item  Estimate one non-Gaussian component from the permuted data. Retain its JB statistic, denoted $f(\hatbS_r^{(t)})$.
    \end{enumerate}
    \item Generate a p-value as $p_r = \frac{1}T \sum_{t=1}^T \left\{ f(\hatbS_r) < f(\hatbS_r^{(t)} \right\}$. 
\end{enumerate}
One can use a binary search algorithm starting from $\lceil n/2 \rceil$ components. If $p_{\lceil n/2 \rceil} >\alpha$, we test $\lceil n/4 \rceil$ components; if $p_{\lceil n/2 \rceil}<\alpha$, we test $\lceil 3n/4 \rceil$ components, as described in \citep{zinlngca}. This reduces the number of tests to $\lceil \log_2 n \rceil$. %This motivates a priori defining $\alpha = 0.05/\lceil \log_2 n \rceil$. 

We evaluate the estimation performance of the proposed scheme for selecting the number of components. We consider $\alpha \in \{0.01, 0.05, 0.1\}$ for the identification of the total and joint number of non-Gaussian components. We also consider the saturated model ($r_x = r_y = 48$) with the number of joint components $r_J$ identified according to Section~\ref{sec:rJ} with $\alpha = 0.01$. 

\begin{figure}[!t]
    \centering
    \includegraphics[width=0.8\textwidth]{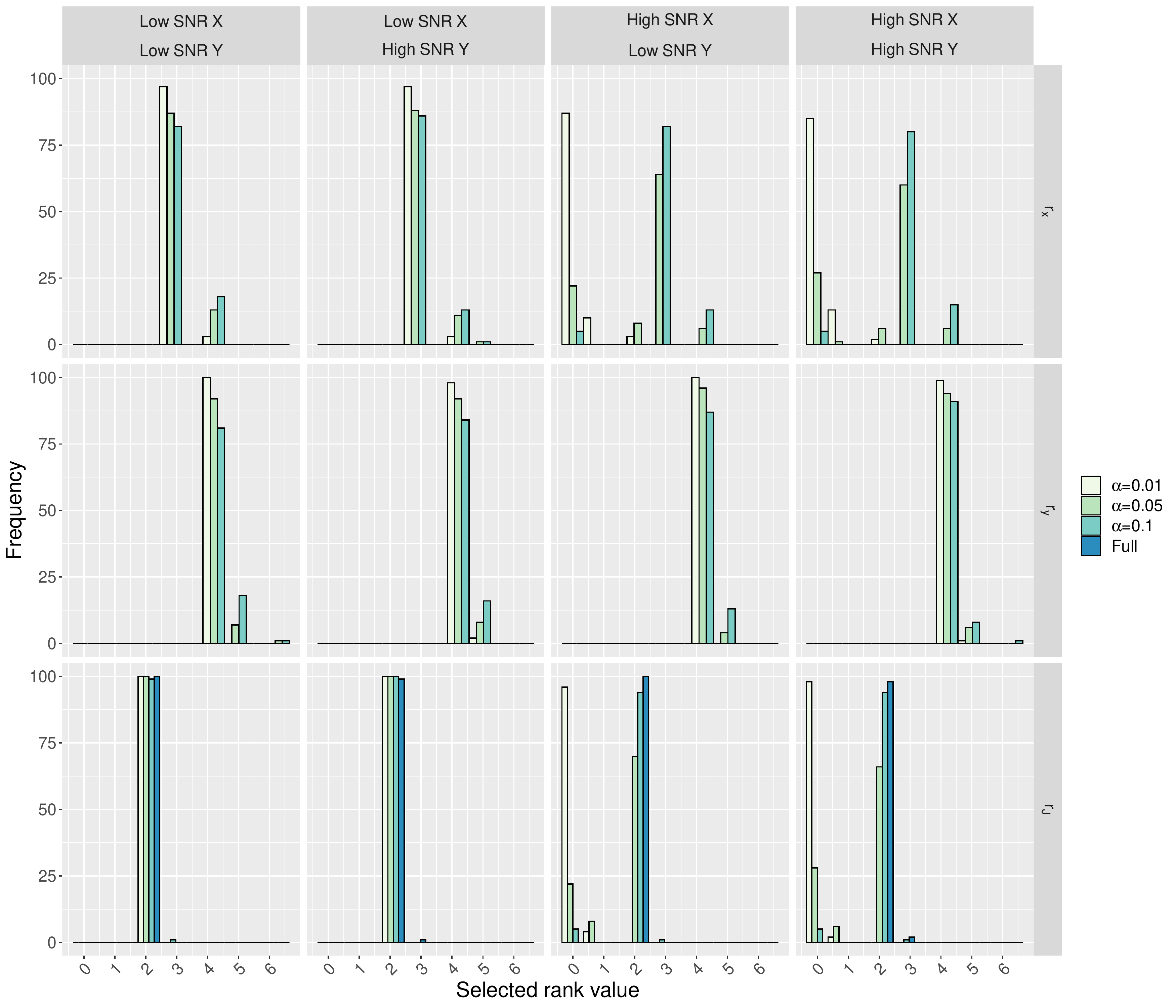}
    \caption{Simulation Setting 1 of Section~\ref{sec:sim_small}: selection frequencies over 100 simulations for each combination of signal to noise ratios (SNR). The true values are $r_x = 3$, $r_y = 4$, and $r_J = 2$. In the legend, $\alpha=0.01,0.05,0.10$ correspond to the $\alpha$-level used in the estimation of the signal rank ($r_x$ and $r_y$). In selecting $r_J$, ``Full'' corresponds to estimating $r_J$ when setting $r_x=r_y=48$ with $\alpha=0.01$ in the joint-rank test.}
    \label{fig:sim_small_ranks}
\end{figure}

Figure~\ref{fig:sim_small_ranks} shows rank selection results over 100 simulations for each SNR combination. In the low SNR regime, we found that the proposed permutation scheme exhibits excellent performance in selecting the initial number of components $r_x$ and $r_y$, with the best performance achieved by $\alpha = 0.01$. In the high SNR regime for $\bX$, the initial rank $r_x$ is consistently underestimated for $\alpha = 0.01$, with the best performance achieved by $\alpha = 0.1$. We conjecture that this is because $p_x$ is much smaller than $p_y$, and therefore the random permutation of the columns is not sufficient in eliminating non-Gaussianity in the high SNR regime thus leading to higher values of JB-statistics on permuted data, and subsequently higher $p$-values. While underestimation of initial ranks $r_x$ and $r_y$ leads to corresponding underestimation of the joint rank $r_J$, the over-estimation of initial ranks has minimal effect on estimation of $r_J$. Specifically, we found that using the saturated model leads to correct $\widehat r_J=2$ in 397 out of 400 simulations. In the remaining 3 simulations, the $r_J$ is slightly over-estimated with $\widehat r_J =3$. Since our focus is on the analysis of joint structure in~\eqref{eq:model}, we use the saturated model in the analysis of the Human Connectome Project data (Section~\ref{sec:data}).

\section{Alternative simultaneous NGCA algorithm based on averaging of mixing matrices}\label{sec:averageSING}

Consider solving two separate NGCA problems (problem~\eqref{eq:joint_app3} with $\rho = 0$). Let the resulting mixing matrices be $\widehat \bM_x$ and $\widehat \bM_y$, and the resulting components be $\widehat \bS_x$ and $\widehat \bS_y$.  Consider matching the columns of $\widehat \bM_x$ and $\widehat \bM_y$ using the greedy algorithm with $r_J$ estimated using the permutation test (Section \ref{sec:rJ}). Denote these permuted estimates with first $r_J$ matched columns as $\widehat \bM_{x}^{J}$ and $\widehat \bM_{y}^{J}$. These matched $\widehat \bM_{x}^{J}$ and $\widehat \bM_{y}^{J}$ are used to initialize the curvilinear search algorithm for~\eqref{eq:joint_app3} as described in Section~\ref{appendix:InitialAlgorithm}. One disadvantage of formulation~\eqref{eq:joint_app3}, however, is that the solution in general depends on the choice of $\rho$.

An alternative approach is to use the matched $\widehat \bM_{x}^{J}$ and $\widehat \bM_{y}^{J}$ directly to construct a common $\widehat \bM_J$ by averaging. Let $\widehat \bM_x^J = \widetilde \bM_{x}^J\bD_x$ and $\widehat \bM_y^J = \widetilde \bM_{y}^J\bD_y$, where the columns of $\widetilde \bM_{x}^J$ and $\widetilde \bM_y^J$ have Euclidean norm one, and the matched column pairs have positive inner product (thus, $\bD_x$ and $\bD_y$ are diagonal matrices capturing the original column scaling and sign invariance). Then the estimator of the common mixing matrix $\widehat \bM_J$ is obtained by taking the average of the matched directions in $\widetilde \bM_{x}^J$ and $\widetilde \bM_{y}^{J}$, with subsequent adjustment of scaling to have Euclidean norm one. That is, the $r$th column of $\widehat \bM_J$ satisfies
$$
\widehat m_{r}^J \propto (\widetilde m_{xr}^J + \widetilde m_{yr}^J)/2.
$$

One of the disadvantages of averaging is that the matrix decomposition~\eqref{eq:modellngca} no longer holds for the new averaged $\widehat \bM_J$ because $\widehat \bM_J^{-} \bX_c$ does not result in a matrix with orthogonal rows. One needs to re-estimate the scaling matrices $\bD_x$, $\bD_y$ as well as components $\bS_{Jx}$, $\bS_{Jy}$. Let $\widehat \bJ_x = \widehat \bM_{x}^{J}\widehat \bS_{x}^J$ and $\widehat \bJ_y = \widehat \bM_{y}^{J}\widehat \bS_{y}^J$ be the original estimated joint signal based on two separate NGCA problems (problem~\eqref{eq:joint_app3} with $\rho = 0$), and let $\widehat \bM_J$ be the estimate of the common mixing matrix obtained by averaging as described above. To re-estimate the components $\bS_{Jx}$, we propose to solve
\begin{align}\label{eq:Sproj}
    \minimize_{\bS_x, \bD_x}\|\widehat \bJ_x - \widehat \bM_J\bD_x\bS_x\|_F^2\quad\mbox{subject to}\quad \bS_x\bS_x^{\top}=p_x\bI,
\end{align}
and similarly for $\bS_{Jy}$. Here the objective function encourages agreement of the signal with the original separate NGCA matrix decomposition, and the constraint ensures the orthogonality of the components. Since all columns of $\widehat \bM_J$ have Euclidean norm one, the scaling adjustment is captured via the diagonal matrix $\bD_x$.

To solve~\eqref{eq:Sproj}, we use alternating minimization with respect to $\bS_x$ and $\bD_x$. Given $\bD_x$, problem~\eqref{eq:Sproj} is an orthogonal Procrustes problem, hence the solution has a closed form based on the singular value decomposition of $\widehat \bJ_x^{\top}\widehat \bM_J\bD_x = \bU\bL\bV^{\top}$, that is $\bS_x = \sqrt{p_x}\bU\bV^{\top}$. Given $\bS_x$, problem~\eqref{eq:Sproj} is quadratic minimization with respect to the diagonal elements of $\bD_x$, which also has a closed form solution $\bD_{x} = \mbox{diag}(\bS_x\widehat\bJ_x^{\top}\widehat \bM_J)$. We alternate the two updates until the change in $|| \bD_x^{k-1} - \bD_x^{k}||_{\infty} < 0.1$. The solution for $\widehat \bS_y$, $\widehat \bD_y$ is similar. We refer to this estimation approach as ``SING-averaged."

\section{Simulation Setting 1}
In all analyses, we use our own implementation of Joint ICA and mCCA+jICA, which allows us to use the JB statistic for all methods. We do not compare to existing toolboxes for multimodal neuroimaging analysis such as `Fusion ICA Toolbox' \citep{rachakonda2012fusion} because (i) they require customization of MATLAB toolboxes \citep{lerman2017multimodal} to make them applicable for surface data and correlation matrices, and (ii) they do not use the JB statistic. 

\begin{figure}[!t]
    \centering
    \includegraphics[width=0.9\textwidth]{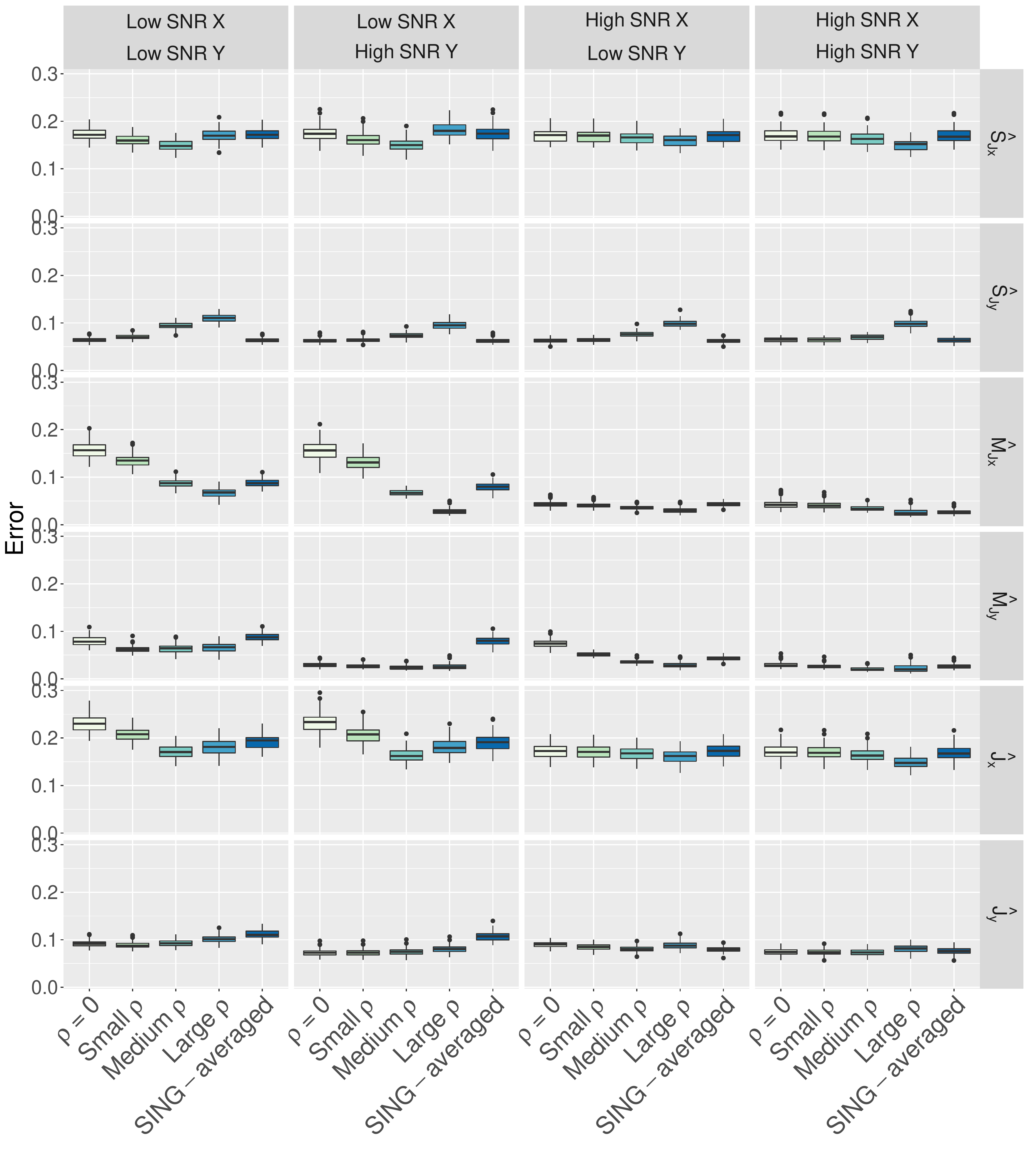}
    \caption{The performance of SING variants in Simulation Setting 1 of Section~\ref{sec:sim_small}. Results are over 100 replications for each combination of signal to noise ratios (SNR). The errors are evaluated using $\sqrt{PMSE}$ from~\eqref{eq:pmse} (components and mixing matrices). By definition, $\sqrt{\text{PMSE}}\in [0, \sqrt{2}] = [0, 1.414]$. The errors on $\widehat \bJ_x$ and $\widehat \bJ_y$ are evaluated using $\sqrt{\text{MSE}}$ from~\eqref{eq:mse}.}
    \label{fig:SINGsetting1}
\end{figure}

Figure \ref{fig:SINGsetting1} displays the accuracy of the joint loadings, joint scores, and the reconstructed joint signals for different values of $\rho$ as well as SING-averaged.

\pagebreak
\clearpage

\begin{table}[!t]
    \centering
    \begin{tabular}{|l|c|c|c|c|c|c|}
    \hline
         Method& Error $\widehat \bS_{Jx}$& Error $\widehat\bS_{Jy}$ & Error $\widehat \bM_{Jx}$ & Error $\widehat \bM_{Jy}$ & Error $\widehat \bJ_x$ & Error $\widehat \bJ_y$ \\
         \hline
         Joint ICA&1.408 & 1.409 & 1.362 & 1.362 & 9.452 & 7.817 \\ 
         mCCA+jICA&1.411 & 1.405 & 1.367 & 1.316 & 9.603 & 7.892 \\  
         $\rho = 0$&0.051 & 0.031 & 0.295 & 0.208 & 0.309 & 0.159 \\ 
         SING-averaged &0.057 & 0.038 & \textbf{0.114} & \textbf{0.114} & 0.202 & 0.175 \\ 
         Small $\rho$&\textbf{0.035} & \textbf{0.023} & 0.118 & 0.121 & 0.134 & 0.101 \\ 
         Medium $\rho$&\textbf{0.035} & \textbf{0.023} & 0.115 & 0.115 & 0.132 & 0.097 \\ 
         Large $\rho$&\textbf{0.035} & \textbf{0.023} & \textbf{0.114} & \textbf{0.114} & \textbf{0.131} & \textbf{0.096} \\
         \hline
    \end{tabular}
    \caption{Simulation setting 2 of Section~\ref{sec:sim_large}. For the components and mixing matrices, $\sqrt{\text{PMSE}}$ values are reported; see ~\eqref{eq:pmse}. By definition, $\sqrt{\text{PMSE}}\in [0, \sqrt{2}]$. The errors on $\widehat \bJ_x$ and $\widehat \bJ_y$ are evaluated using $\sqrt{\text{MSE}}$ from~\eqref{eq:mse}.}   \label{tab:sim_large}
\end{table}

\section{Simulation Setting 2}
In simulation setting 2, both Joint ICA and mCCA+jICA perform poorly compared to the SING variants, both with respect to the components and the scores (Table~\ref{tab:sim_large}). The separate estimation approach ($\rho = 0$) leads to accurate $\hatbS_{Jx}$ and $\hatbS_{Jy}$ but poorer estimates of $\hatbM_{Jx}$ and $\hatbM_{Jy}$ relative to $\rho>0$ and averaging, while the averaging approach leads to accurate $\hatbM_{Jx}=\hatbM_{Jy}$ but comparatively less accurate $\hatbS_{Jy}$ and $\hatbS_{Jx}$. Results are similar across small, medium, and large $\rho$, as the $\rho$s evaluated here all resulted in $\hatbM_{Jx} \approx \hatbM_{Jy}$.

\begin{figure}
    \centering
    \includegraphics[width=\textwidth]{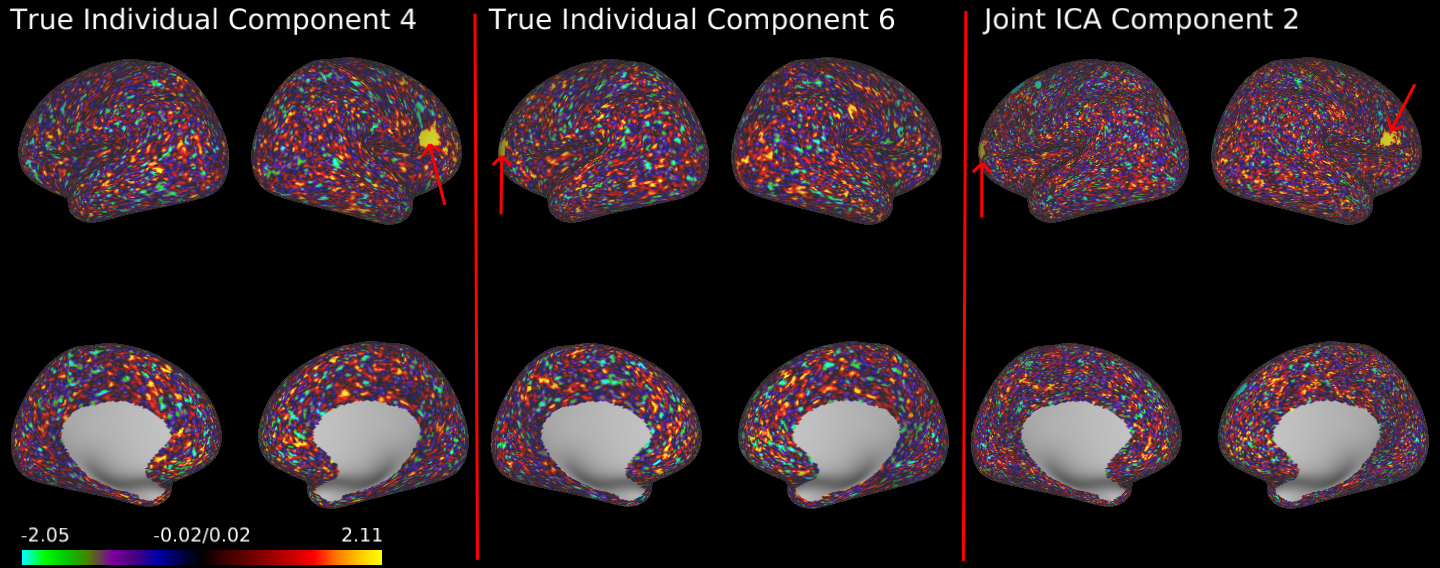}

    \caption{Simulation setting 2: Joint ICA and mCCA+jICA component 2. The Joint ICA component contains features from true individual components 4 and 6 rather than information from the joint structure. The mCCA+jICA contains features from the true individual component 4.}\label{fig:simspur2}
\end{figure}

Figure \ref{fig:simspur2} depicts the second joint component estimated by Joint ICA and mCCA+jICA. In both methods, large features loadings from the individual signal contaminated the joint signal. Joint ICA component 2 was contaminated by the true individual component 4 and true individual component 6. Component 2 in mCCA+jICA was contamined by the true individual component 6.

\begin{figure}[!t]
    \centering
    \includegraphics[width =\textwidth]{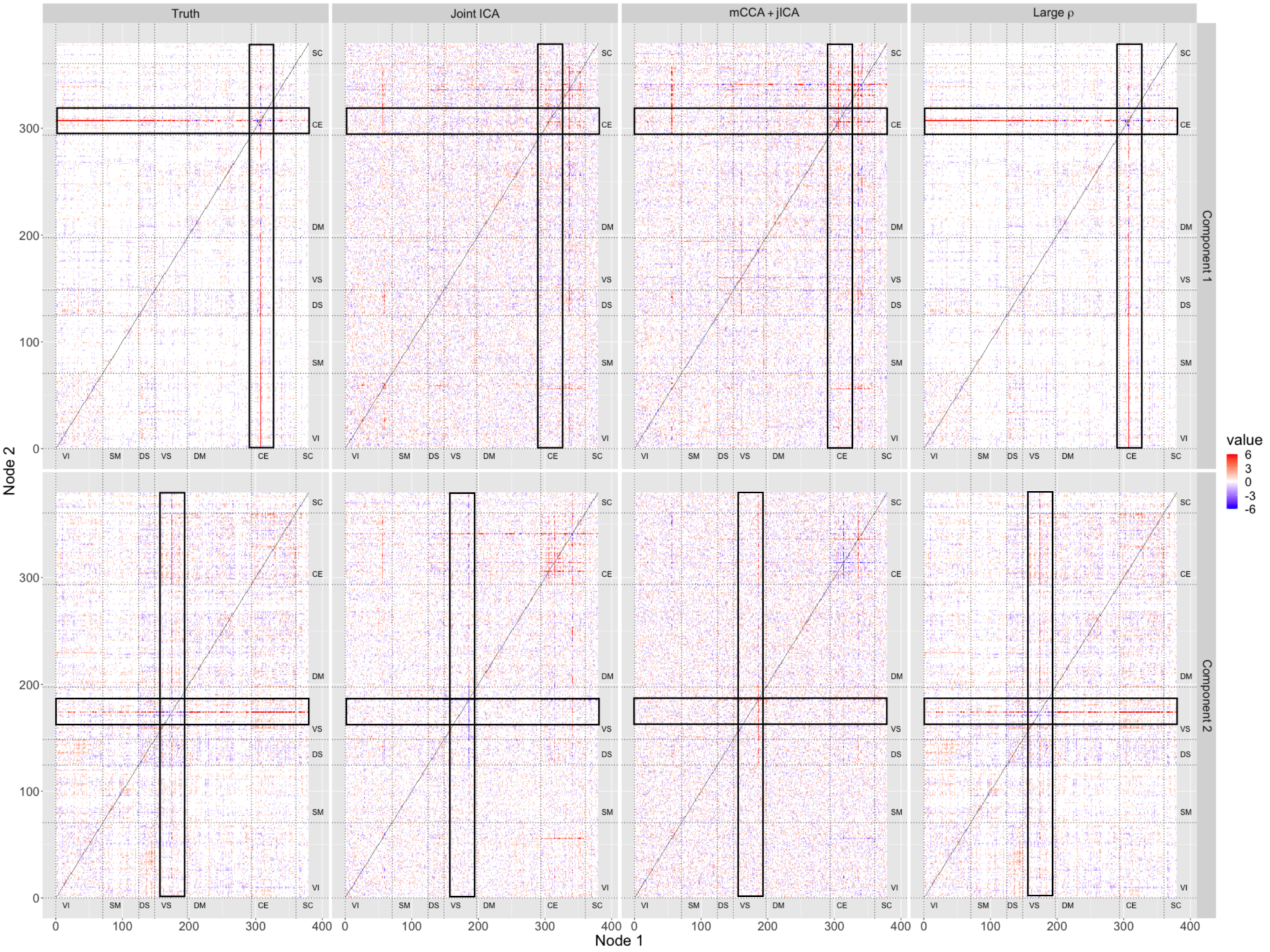}\\
    \includegraphics[width = \textwidth]{LargeScaleBothComponentsY_mCCA.pdf}
    \caption{Simulation setting 2 of Section~\ref{sec:sim_large}, comparison of loading values of the joint component of $\bY$. Large $\rho$ refers to the SING algorithm applied to joint components with $\rho$ equal ten times the value of JB statistic. Top: comparison of estimated network loadings; Bottom:~comparison of L1 norm values of the rows in the estimated network loadings.}
    \label{fig:Ycomponent1}
\end{figure}
\pagebreak
\clearpage

\section{Simulation Setting 3}\label{sec:sim_sparse}

In this section, we compare the performance of the methods in the large-scale simulation study when the true components are exactly sparse. Specifically, we modify the true components in Simulation Setting 2 (Section~\ref{sec:sim_large}) by setting to zero all elements below 5 in absolute value. This leads to 99.9\% zeros for each component for $\bX$, and 99.8\% zeros for $\bY$. As truncation induces violations of condition 3 for model~\eqref{eq:model} (mean zero, uncorrelated, diagonal scaling), we further re-normalize the truncated components to satisfy this condition and at the same time preserve the sparsity pattern. The resulting new sparse components are quite different from the original ones, with $\sqrt{\text{PMSE}} = 1.25$ between the joint components from Setting 2 and joint components from Setting 3 for $\bX$, and $\sqrt{\text{PMSE}} = 1.14$ for $\bY$.

\begin{table}[!t]
\centering
 \begin{tabular}{|l|c|c|c|c|c|c|}
  \hline
 Method& Error $\widehat \bS_{Jx}$& Error $\widehat\bS_{Jy}$ & Error $\widehat \bM_{Jx}$ & Error $\widehat \bM_{Jy}$ & Error $\widehat \bJ_x$ & Error $\widehat \bJ_y$ \\
  \hline
Joint ICA & 1.407 & 1.409 & 1.363 & 1.363 & 9.477 & 7.808 \\ 
  mCCA+jICA & 1.408 & 1.406 & 1.365 & 1.312 & 9.641 & 7.901 \\ 
  $\rho = 0$ & 0.029 & \textbf{0.015} & 0.225 & 0.168 & 0.229 & 0.115 \\ 
  SING-averaged & 0.051 & 0.036 & 0.127 & 0.127 & 0.155 & 0.147 \\ 
  Small $\rho$& \textbf{0.023} & 0.023 & 0.095 & 0.084 & 0.101 & 0.079 \\ 
  Medium $\rho$ & 0.024 & 0.026 & 0.080 & \textbf{0.079} & 0.088 & \textbf{0.074} \\ 
  Large $\rho$& 0.024 & 0.026 & \textbf{0.079} & \textbf{0.079} & \textbf{0.087} & \textbf{0.074} \\ 
   \hline
\end{tabular}
    \caption{Simulation Setting 3 (true components are sparse). Reported are values of $\sqrt{\text{PMSE}}$ - the root of permutation- and scale-invariant mean squared error from~\eqref{eq:pmse} (components and mixing matrices). By definition, $\sqrt{\text{PMSE}}\in [0, \sqrt{2}] = [0, 1.414]$. The errors on $\widehat \bJ_x$ and $\widehat \bJ_y$ are evaluated using $\sqrt{\text{MSE}}$ from~\eqref{eq:mse}.}
    \label{tab:sim_sparse}
\end{table}

Table~\ref{tab:sim_sparse} compares the performance of different methods in this scenario. Despite the exact sparsity of the new components, the conclusions are similar to Simulation settings~1--2. Joint ICA and mCCA+jICA perform worse compared to SING variants. The separate approach ($\rho = 0$) is accurate for $\hatbS_{Jx}$ and $\hatbS_{Jy}$ but not $\hatbM_{Jx}$ and $\hatbM_{Jy}$. Using $\rho > 0$ in SING leads to improvements in $\hatbM_{Jx}$ and $\hatbM_{Jy}$. SING-averaged performs well for $\hatbM_{Jx}$ and $\hatbM_{Jy}$ but again poorly for $\hatbS_{Jx}$ and $\hatbS_{Jy}$. The reconstructed signal $\hatbJ_x$ and $\hatbJ_y$ show marked improvements in $\rho>0$ versus $\rho=0$ and SING-averaged. Finally, while none of the SING variants impose sparsity of the components, the resulting estimation error is very small. Thus, SING can accurately estimate underlying sparse components even though it does not rely on sparsity regularization.

\section{Supplementary Information for the Analysis of the Human Connectome Project Data}\label{appendix:HCP}

We used the statistical parametric maps (SPMs) from a working memory task in which a subject viewed a sequence of images and was instructed to press a clicker if they had seen that image prior to the previous image, which is known as a 2-back task. The subject also performed a 0-back task, in which an initial image was shown, and the subject pressed the clicker if that image was seen anytime during the task. We used the 2-back 0-back contrast as the input data, which is an estimate of the locations that are used for working memory. Specifically, we used the 2-mm smoothed cifti files provided by the HCP of the name {\texttt{<subjectID>\_tfMRI\_WM\_level2\_hp200\_s2\_MSMAll.dscalar.nii}}. See \cite{barch2013function,glasser2013minimal} for details. We extracted the 59,412 cortical surface vertices from each of these task maps.

\begin{figure}[!t]
    \centering
    \includegraphics[width=0.75\textwidth]{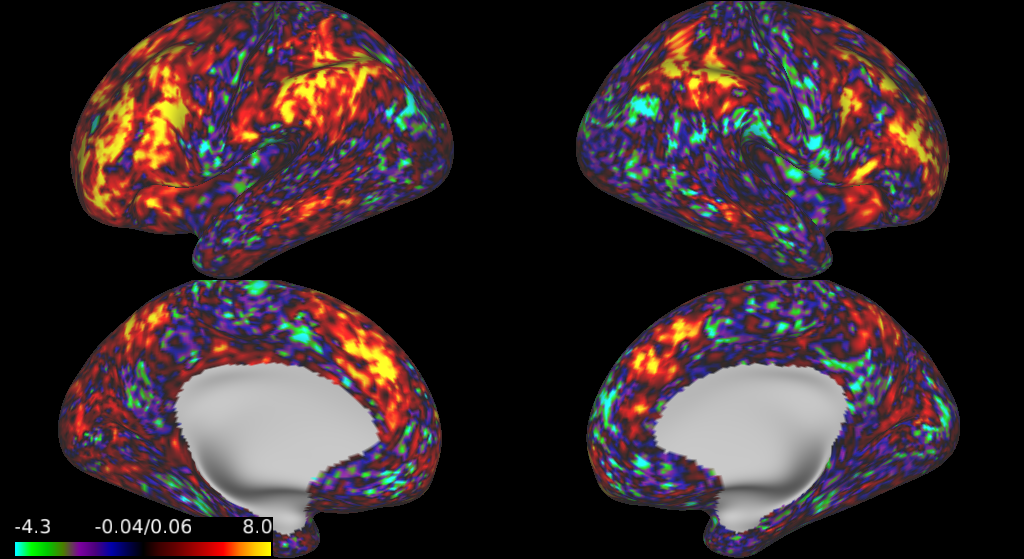}
    \caption{Example subject's data (subject ID 101915). 2bk - 0bk task activation map. The data are vectorized for input to SING, such that each subject's image is a row of $\bX$ of length 59,412.}\label{fig:example}
\end{figure}

\begin{figure}[!t]
\centering
\includegraphics[width=0.75\textwidth]{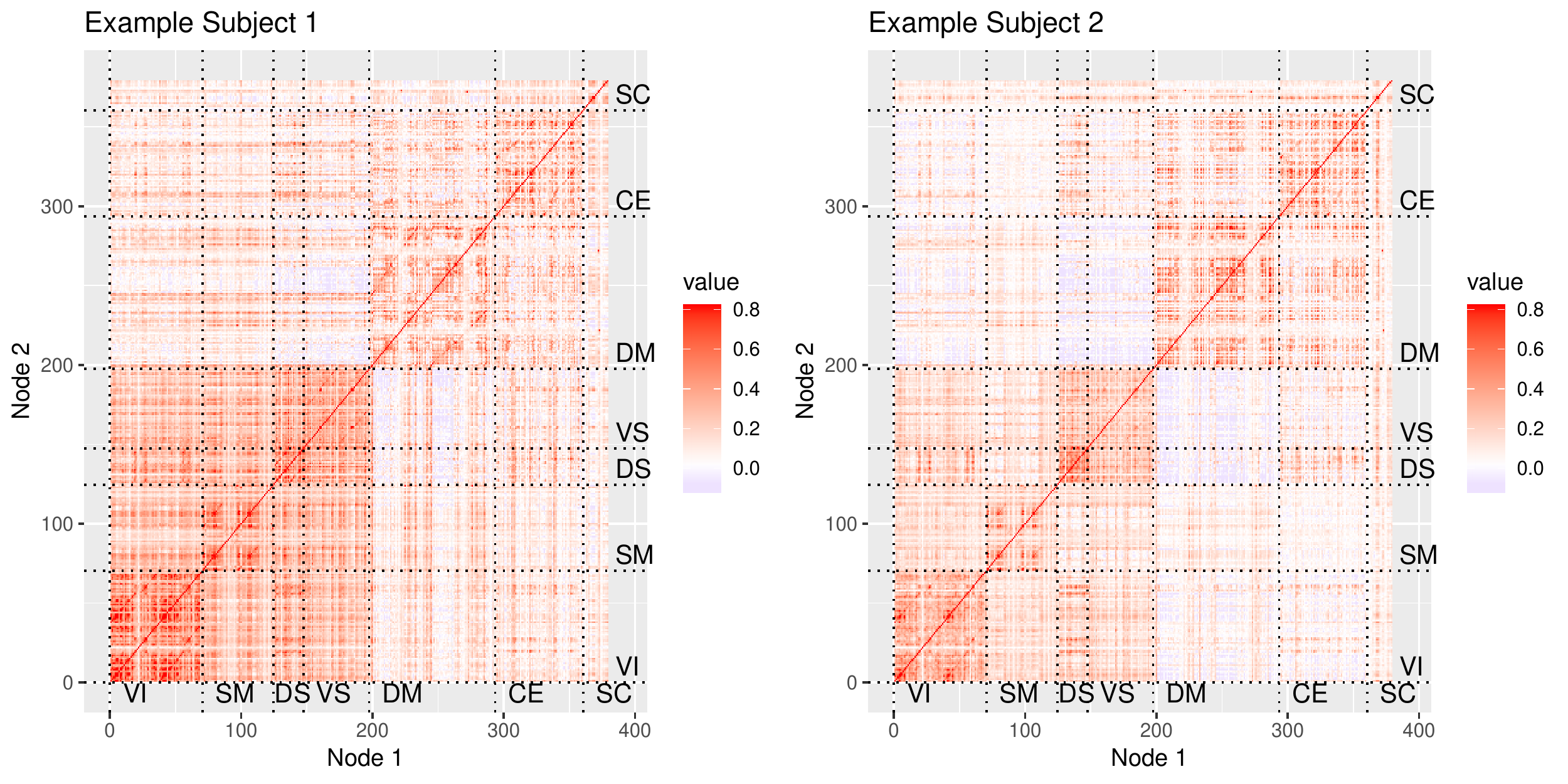}
\caption{Resting-state correlation matrices from two example subjects (subject IDs 100206 and 100307). The lower diagonal elements are vectorized for input to SING, such that each subject's connectivity matrix is a row of $\bY$. The nodes from the multi-modal parcellation \citep{glasser2016multi} have been ordered according to communities from \cite{akiki2018determining}. VI: Visual; SM: somatomotor; DS: dorsal salience; VS: ventral salience; DM: default mode; CE: central executive; SC: subcortical (grayordinate voxels including cerebellum).}
\label{fig:example2}
\end{figure}

The HCP data includes 4 resting-state scans of 14:33 each in which two scans are collected for each of two phase-encoding directions (left-right and right-left). We used the ICA-FIX preprocessed data of the name {\texttt{<subjectID>\_rfMRI\_REST1\_LR\_Atlas\_MSMAll\_hp2000\_clean.dtseries.nii}}; see \cite{smith2013resting} for additional details. For each scan, the time series of each vertex was centered and normalized to unit variance. Then we averaged surface vertices according to regions defined by the multi-modal parcellation (360 regions) \citep{glasser2016multi} plus the subcortical gray matter structures (19 regions), resulting in 379 time series for each scan. The correlations were then calculated and Fisher-transformed for each scan, and then the four scans were averaged. 

\begin{figure}[!t]
    \centering
    \includegraphics[width = 0.6\textwidth]{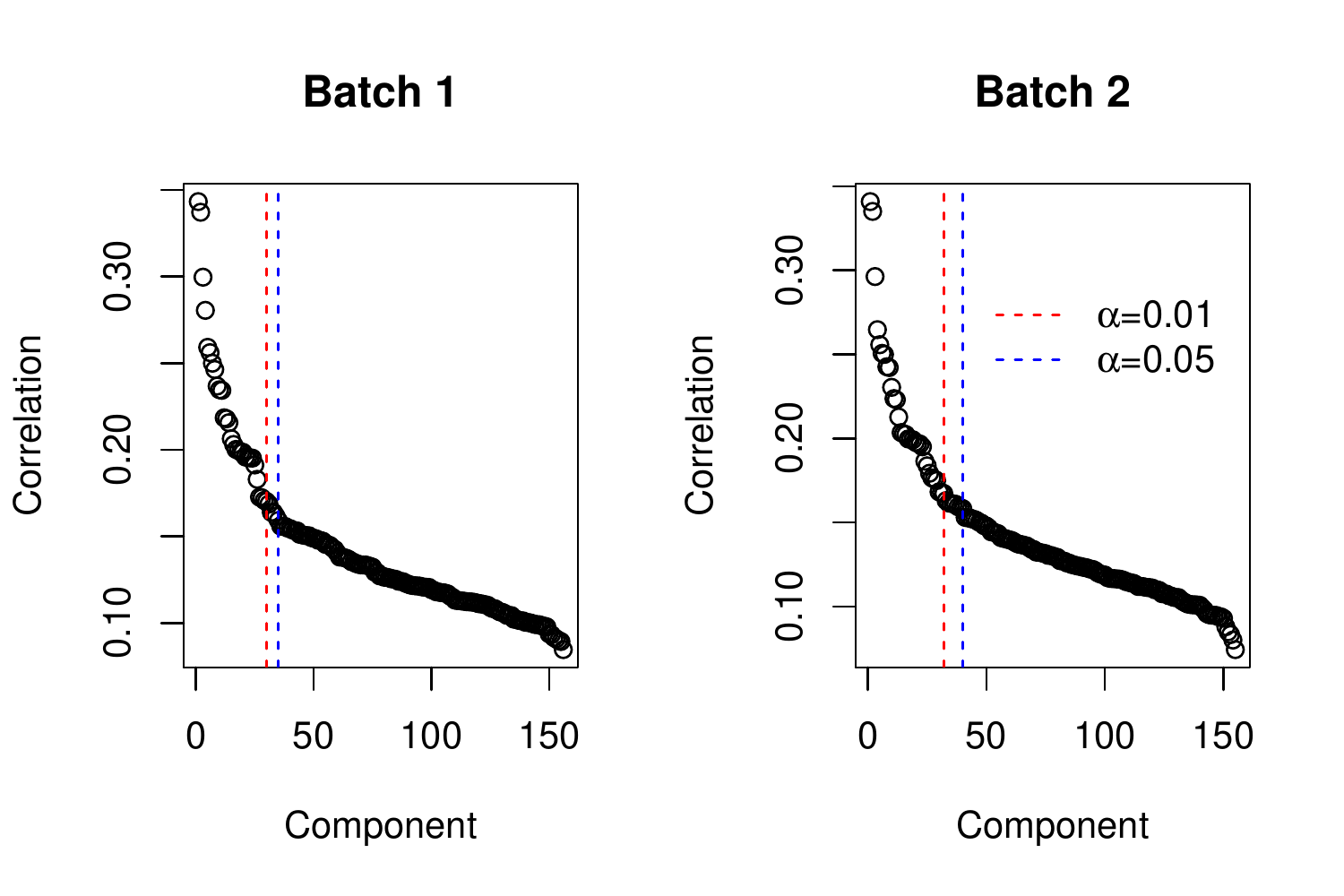}
    \caption{Plots of the matched correlations from the permutation test of the joint rank for the task (working memory) maps and rs correlations. Components from $\alpha=0.01$ that were consistent across the two batches (correlation between  loadings$>$0.95) were retained for input to SING.}
    \label{fig:hcp_jointrank}
\end{figure}

\begin{figure}[!t]
    \centering
    \includegraphics[width=\textwidth]{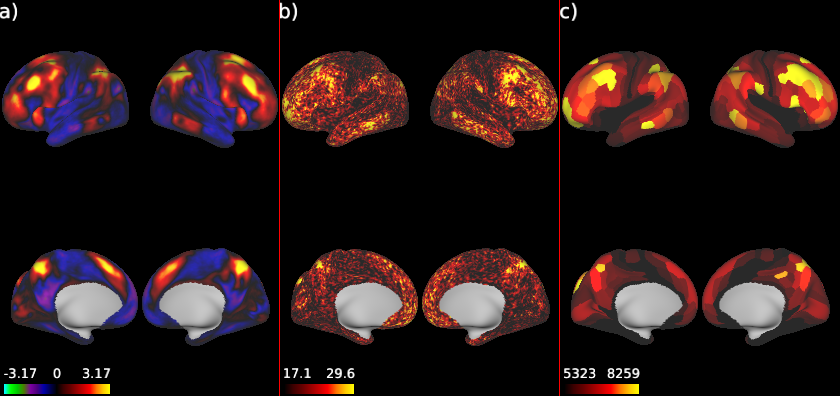}
    \caption{A) The average of the subject 2 bk - 0 bk task (working memory) maps. B) Composite image of the task map loadings formed by summing the absolute value of the task map loadings across components. C) Composite image of the rs correlation loadings formed by summing the absolute values of the edges for each node (L1 norms of the edges), then summing across components. Panel B) shares some of the same features as Panel A), but also differences. The mean is removed during preprocessing, and similarities reflect a relationship between higher order moments (maximized by SING) and the mean. Panel C) generally has higher values in regions that correspond to vertices with higher values in Panel B). The similarities between Panels B) and C) reflect the general spatial correspondence of the joint components.}
    \label{fig:composite}
\end{figure}

\begin{figure}[!t]
    \centering
    \includegraphics[width=\textwidth]{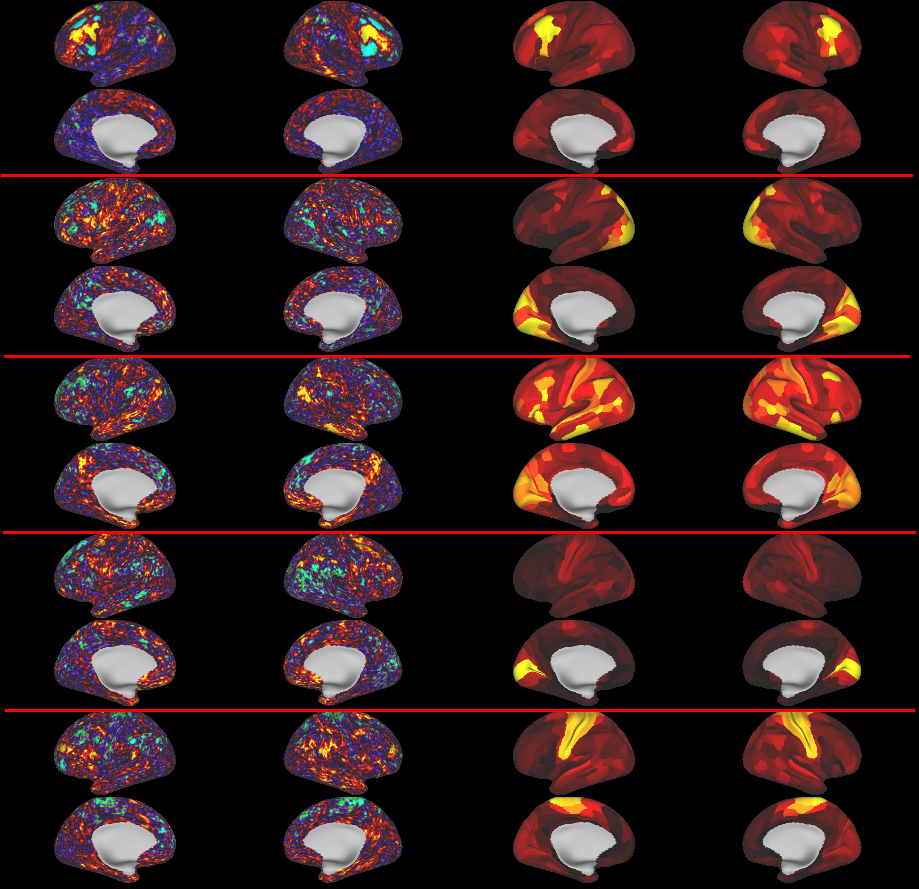}
    \caption{The first five components from Joint ICA. Each component has four views (clockwise: lateral left, lateral right, medial right, medial left) from the task map component (left two columns) and four views from the L1-norms of the edges for each node in the symmetric loadings matrix from the rs correlation component (right two columns). The first component appears to contain joint structure in the working memory and resting-state data sets, but the joint structure in the other components is unclear. In all images, the colorbar is set with absolute percentage [2, 99.9].} \label{fig:jointICA5}
\end{figure}

\begin{figure}[!t]
    \centering
    \includegraphics[width=\textwidth]{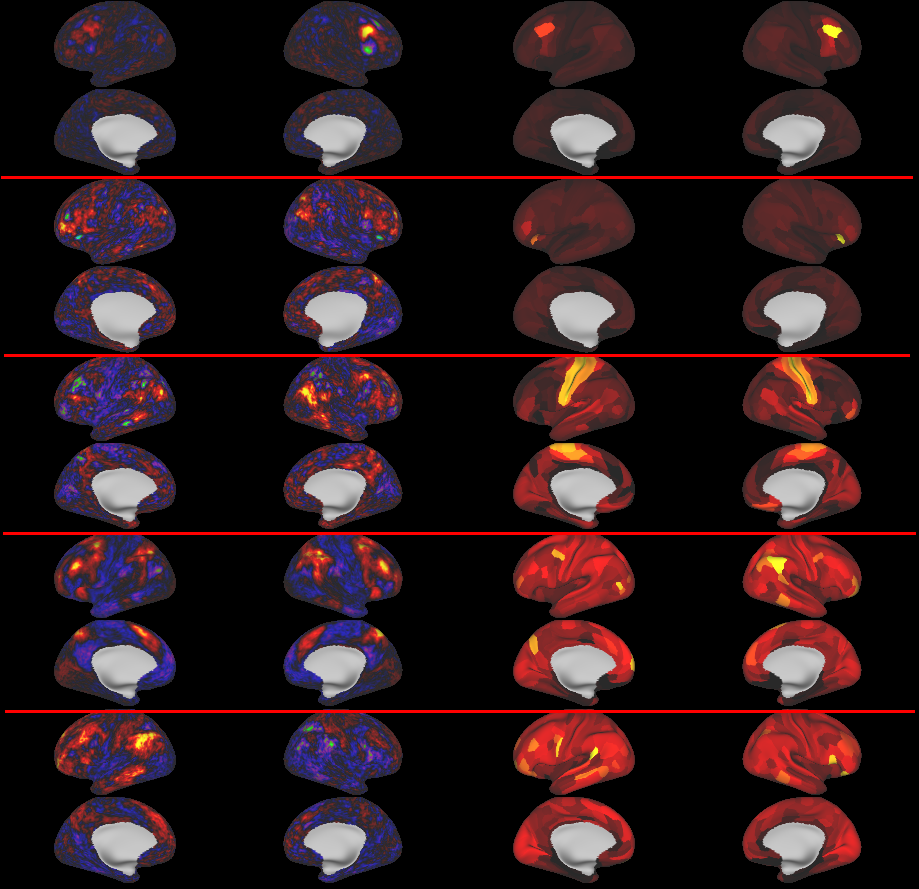}
    \caption{Components 1, 7, 13, 19, and 15 from mCCA+jICA. Each component has four views (clockwise: lateral left, lateral right, medial right, medial left) from the task map component (left two columns) and four views from the L1-norms of the edges for each node in the symmetric loadings matrix from the rs correlation component (right two columns). In all images, the colorbar is set with absolute percentage [2, 99.9].} \label{fig:mCCA5}
\end{figure}

In the multiple regression with fluid intelligence, there are three additional components with $p<0.05/26$ in SING.  In Joint ICA, there were six additional components with $p<0.05/26$. In mCCA+jICA, there were three additional components with $p<0.05/26$ in the working memory scores and of these, two also had $p<0.05/26$ in their corresponding rs correlation scores. 

\begin{figure}[!h]
    \centering
    \includegraphics[width=\textwidth]{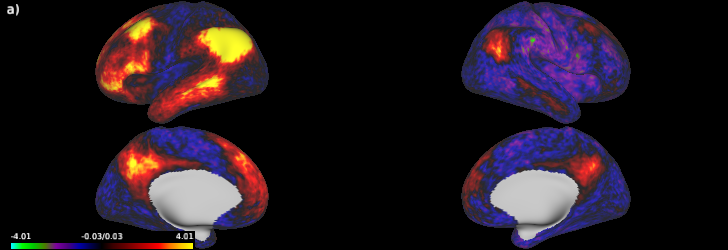}
    \includegraphics[width=0.32\textwidth]{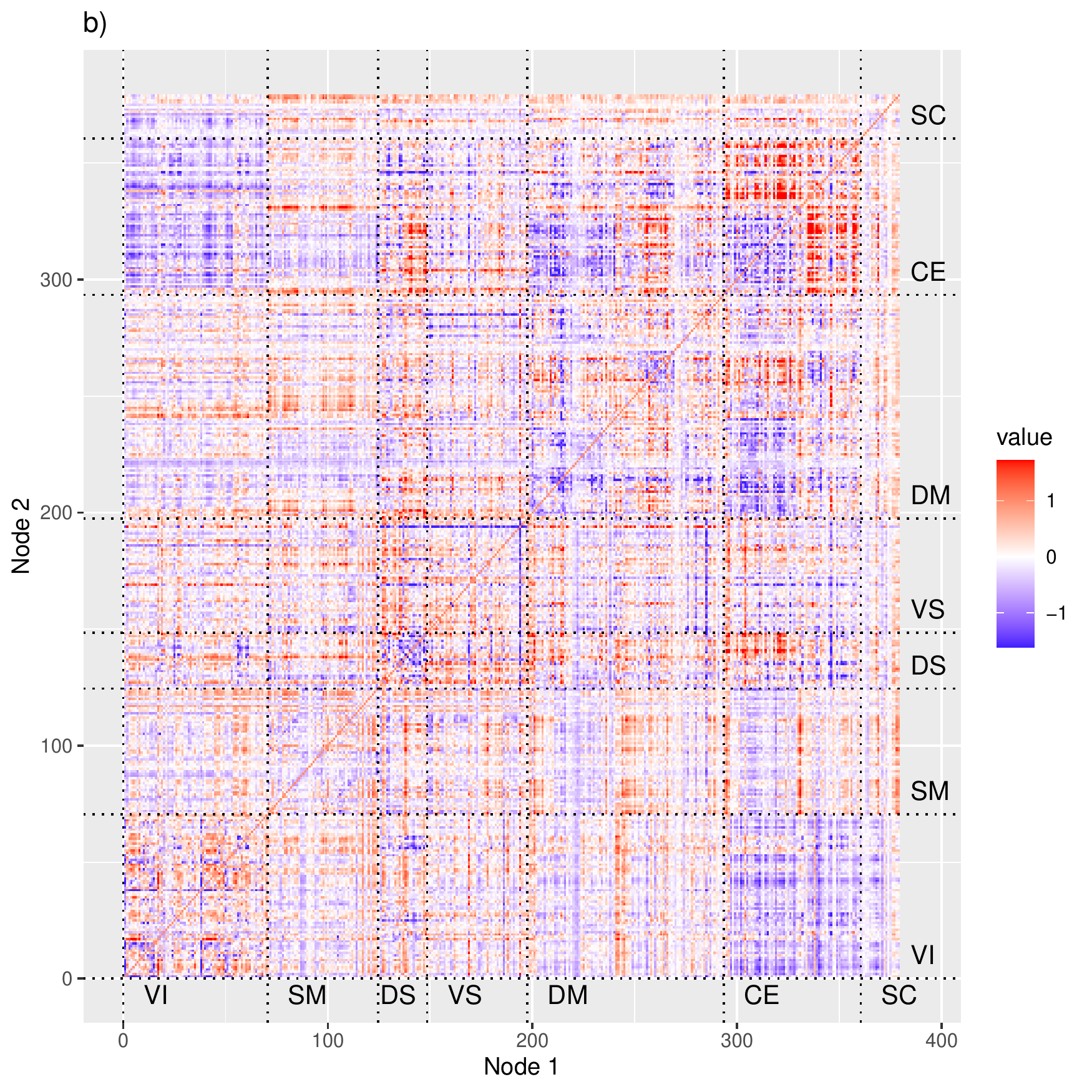}
    \includegraphics[width=0.32\textwidth]{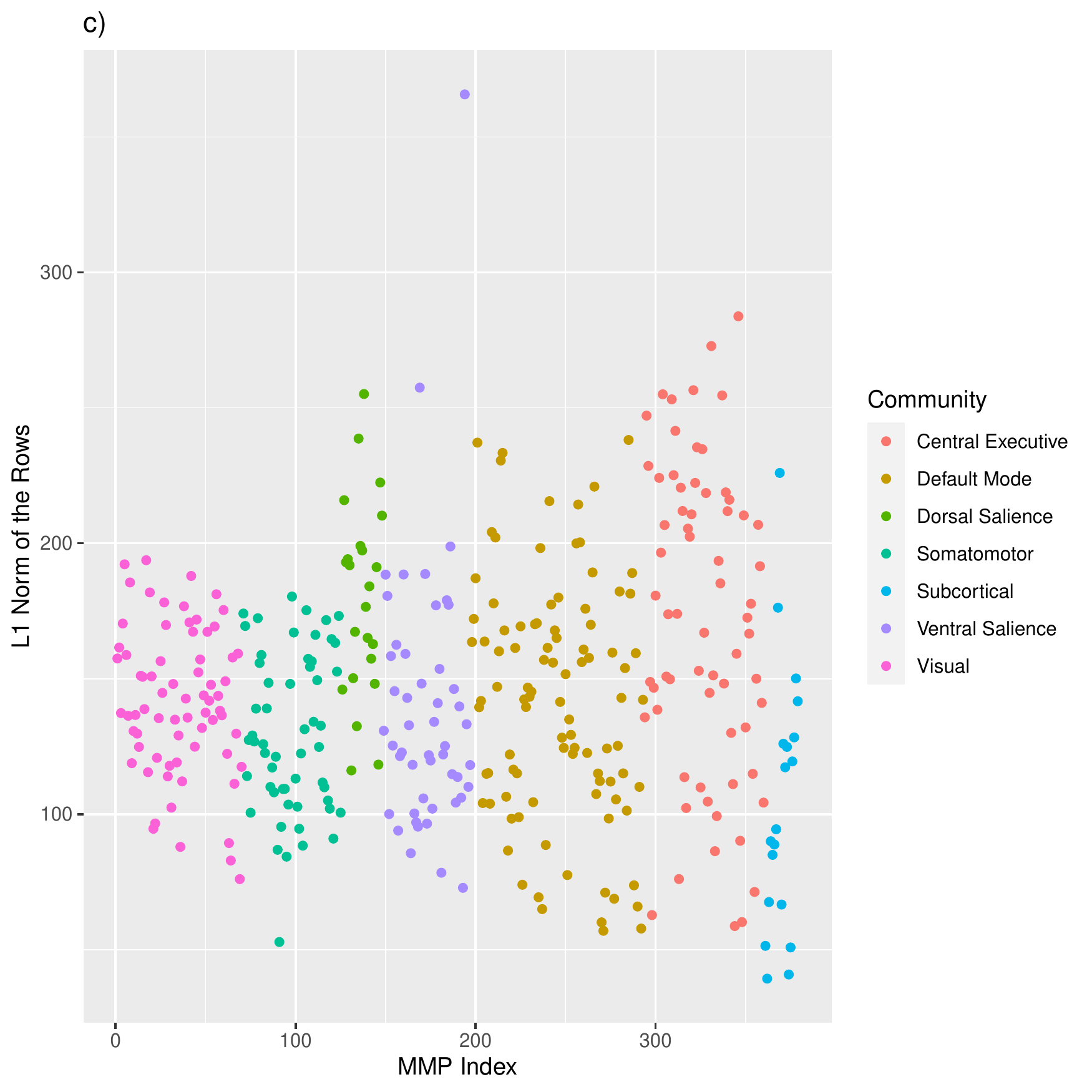}
    \includegraphics[width=0.32\textwidth]{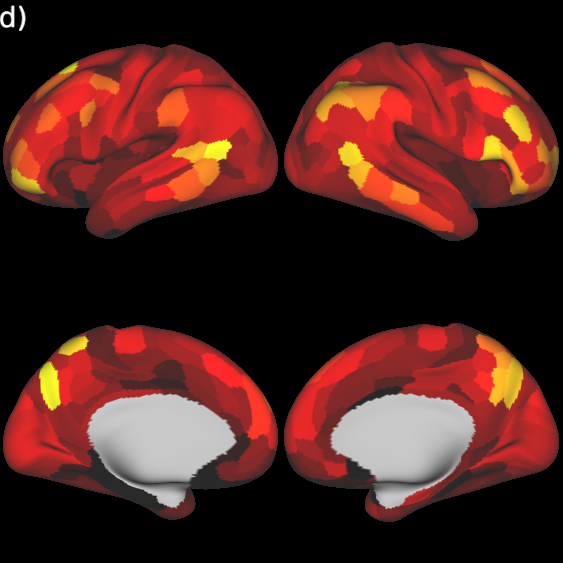}
    \caption{Joint ICA component corresponding to subject scores most strongly related to fluid intelligence ($t=-6.41$, $p<1e-09$). In this component, the task map loadings contain 85\% of the component variance, suggesting the task map may be contributing to the subject scores more than the rs correlations. The component from the task map highlights parts of the default mode network, which do not appear to be prominent in the rs correlations.}
    \label{fig:hcpjointica}
\end{figure}

We provide a detailed examination of the component from Joint ICA with scores most strongly related to fluid intelligence  ($t=-6.41$, $p<1-09$). In Figure \ref{fig:hcpjointica}, the task map component contains loadings from large areas of cortex including parts of the left hemisphere of the default mode network. This suggests that subjects with high scores for this component have lower fluid intelligence, which is associated with higher activity in the default mode network. The matrix view of the loadings in the rs correlations show some positive loadings between intra-community edges of the central executive, negative loadings between central executive and visual, and negative loadings between central executive and default mode network. These loadings are distributed over many edges and nodes, and there is not a clear spatial correspondence in the joint loadings of the two datasets. 

\end{appendix}

\clearpage

\bibliographystyle{agsm} 
\bibliography{MasterBibliography,IrinaReferences}
\end{document}

%% file: Notation.tex
\def\R{\mathbb{R}}
\newcommand{\tR}{\mathbb{R}} % Ben's habit for the reals
\def\ba{\boldsymbol{a}}

\def\bm{\boldsymbol{m}}
\def\bn{\boldsymbol{n}}

\def\bs{\boldsymbol{s}}

\def\bu{\boldsymbol{u}}

\def\bx{\boldsymbol{x}}
\def\by{\boldsymbol{y}}

\def\bA{\boldsymbol{A}}

\def\bD{\boldsymbol{D}}

\def\bI{\boldsymbol{I}}
\def\bJ{\boldsymbol{J}}

\def\bL{\boldsymbol{L}}
\def\bM{\boldsymbol{M}}
\def\bN{\boldsymbol{N}}

\def\bP{\boldsymbol{P}}

\def\bS{\boldsymbol{S}}
\def\hatbS{\widehat{\bS}}

\def\bU{\boldsymbol{U}}
\def\bV{\boldsymbol{V}}
\def\bW{\boldsymbol{W}}
\def\bX{\boldsymbol{X}}
\def\bY{\boldsymbol{Y}}

\def\ones{\boldsymbol{1}}

\def\bmu{\boldsymbol{\mu}}

\def\bLambda{\boldsymbol{\Lambda}}

\def\bSigma{\boldsymbol{\Sigma}}

\DeclareMathOperator*{\minimize}{minimize}

\DeclareMathOperator*{\argmin}{argmin}

\newcommand{\bzero}{\boldsymbol{0}}
\newcommand{\bone}{\boldsymbol{1}}
\newcommand{\cO}{\mathcal{O}}
\newcommand{\hatbM}{\widehat{\bM}} 
\newcommand{\hatbJ}{\widehat{\bJ}} 

\newcommand{\hatbL}{\widehat{\bL}}
\newcommand{\hL}{\widehat{\bL}}

\newcommand{\iv}{{1\hspace{-4pt}1}}